**Sensitivity of ECMWF's seasonal forecast model to $CO_2$ and anthropogenic aerosol forcings: Experimental design and impact on climate trends**

Michael Mayer[1,2,*], Daniel J. Befort[1], Jacob Maddison[1], and Antje Weisheimer[1,3]

[1] ECMWF, Reading, UK, and Bonn, Germany
[2] Department of Meteorology and Geophysics, University of Vienna, Vienna, Austria
[3] NCAS, University of Oxford, Atmospheric, Oceanic and Planetary Physics, Oxford, UK
[*] Corresponding author, Email: michael.mayer@ecmwf.int

Short title: Seasonal Forecast Sensitivity to $CO_2$ and Aerosol Forcing

**Abstract**

Detection and attribution studies typically rely on free-running climate model experiments with modified anthropogenic forcings, yet they fail to reproduce key observed decadal trends that include a forced component. Despite reduced mean-state biases, higher predictive skill, large ensemble sizes and improved representation of trends through frequent re-initialization, seasonal forecasting models have not previously been used to address these shortcomings.

We introduce a novel set of counterfactual seasonal hindcasts using ECMWF's coupled seasonal forecasting system. These are initialized under alternative forcing scenarios by modifying atmospheric forcing and ocean and sea-ice initial conditions. An observation-based estimate of the forced ocean temperature signal is derived to either amplify or remove this signal from the ocean initial conditions. Retrospective forecasts for 1993–2023 are performed using a control configuration together with enhanced- and reduced-forcing experiments, in which either post-1993 annual $CO_2$ increases and anthropogenic ocean warming are doubled or $CO_2$ is fixed to 1993 levels and the post-1993 anthropogenic ocean-warming signal is removed. Additional experiments isolate the influence of aerosol forcing.

The counterfactual hindcasts substantially alter long-term temperature trends while largely preserving seasonal prediction skill, interannual variability and model drift, demonstrating dynamically consistent perturbations. Enhanced forcing strengthens several observed climate trends underestimated by the control, including top-of-atmosphere radiative fluxes and aspects of the atmospheric circulation over the tropical Pacific. However, strengthening the tropical Pacific zonal sea-surface-temperature gradient produces only a weak atmospheric response. As a result, the coupled atmosphere–ocean system fails to sustain the temperature gradient anomaly through Bjerknes feedbacks, suggesting a fundamental limitation of the model. Amplified aerosol forcing has only a weak impact, likely because indirect aerosol effects are omitted in the model. These results establish counterfactual seasonal hindcasts as a practical framework for dynamical attribution and a powerful tool for diagnosing model deficiencies in responses to anthropogenic climate forcing.

## 1. Introduction

Despite broad agreement between climate models and observations on the evolution of global mean surface temperature over the past ~170 years, the most recent years have exhibited an accelerated warming that has challenged current understanding of the climate system (Schmidt 2024; Minobe et al. 2025; Allan and Merchant 2025). This enhanced warming coincides with the emergence of several discrepancies between historical climate model simulations and observations in both large-scale and regional trends of key climate indices (Simpson et al. 2025; Byrne et al. 2025), including trends associated with the El Niño-Southern Oscillation (ENSO; Zhang et al. 2025). Collectively, these inconsistencies highlight critical knowledge gaps in our understanding of regional climate change and raise questions about the representation of anthropogenic forcings, most notably greenhouse gases and aerosols, in forecasts and projections of future regional climate.

Similar deficiencies have been identified in global coupled forecasting systems used for seasonal prediction. These systems are initialised from the observed state of the climate system and exhibit biases in simulated climate trends, including trends in zonal sea-surface temperature (SST) gradients across the tropical Pacific and seasonal-mean large-scale warming patterns (Beverley et al. 2024; Patterson et al. 2025; Mayer et al. 2025). Seasonal forecasting systems generally employ external forcings similar to those used in the Coupled Model Intercomparison Project (CMIP), while typically operating at higher horizontal and vertical resolution. They are routinely run as large ensembles at forecasting centres worldwide to provide probabilistic predictions on seasonal-to-annual timescales.

Importantly, many of the trend discrepancies identified in historical climate model simulations emerge rapidly in initialized forecasts and can become detectable within months, and in some cases even hours, after initialisation (Mayer et al. 2025). The short time scales over which these errors develop provide a unique opportunity to use initialized forecast systems as tools for diagnosing the origins of model-observation trend discrepancies at a substantially lower computational cost than century-long climate integrations (Simpson et al. 2025).
Initialised seasonal forecasts also offer conceptual advantages over free-running climate simulations for the investigation of climate trends. Through regular re-initialization from observations, forecast ensembles remain closely constrained by the observed evolution of the climate system. Consequently, each ensemble member represents a plausible realization of seasonal weather and climate variability consistent with the observed large-scale state at initialisation. For example, El Niño and La Niña events occur in the same years as observed in seasonal hindcasts, whereas their timing is unconstrained in free-running simulations. This feature may provide advantages for attribution studies of individual events and their associated circulation anomalies.

Furthermore, frequent re-initialization reduces systematic biases and trend errors relative to free-running climate simulations (Fig. 1). As a result, simulated circulation trends evolve from a more realistic background state, increasing confidence in their representation and interpretation (He and Soden 2016 and references therein).

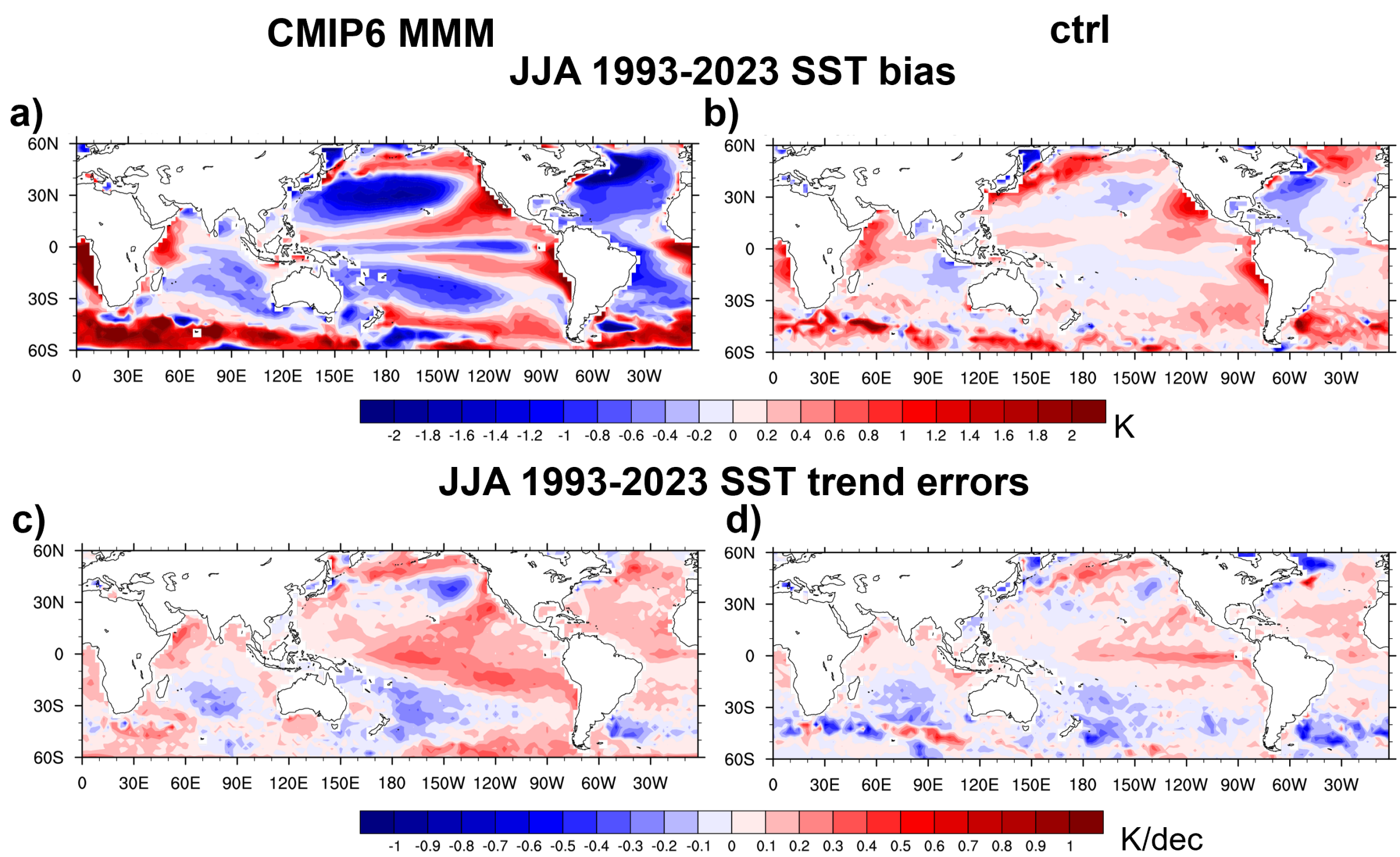

**Figure 1. Comparison of (left) CMIP6 multi-model-mean (MMM) SST (a) bias and (c) trend errors with (right) seasonal hindcasts SST (b) bias and (d) trend error for JJA 1993-2023. The CMIP6 MMM consists of 31 historical simulations combined with SSP585 projections to cover the period. The seasonal hindcast results are taken from the control simulation (CTRL) described in section 2 and are based on 1st May initializations 1993-2023. ERA5 SSTs are used as a reference.**

Established attribution frameworks for free-runing climate simulations include the Detection and Attribution Model Intercomparison Project (DAMIP; Gillett et al. 2016) and the Large Ensemble Single Forcing Model Intercomparison Project (LESFMIP; Smith et al. 2022). These initiatives quantify the contribution of individual forcings to historical climate change through ensembles of coupled climate simulations in which either a single forcing evolves historically while all others remain fixed (single-forcing experiments), or all forcings evolve except one (“all but one” experiments).

In this study, we adapt the underlying principles of these attribution frameworks in the context of initialised seasonal foresting systems. Using ECMWF's state-of-the-art seasonal forecasting system, we investigate the sensitivity of simulated climate trends to changes in radiative forcing from atmospheric $CO_2$ and anthropogenic aerosols. This setup provides a unique framework for examining the development of historical climate trends in initialised hindcasts and assessing their dependence on the imposed radiative forcing.
To this end, we perform a suite of hindcast experiments with modified radiative forcing and consistently adjusted ocean initial conditions following the approaches proposed by Leach et al. (2024) and Weisheimer et al. (2025). By perturbing both the atmospheric forcing and the corresponding forced component of the ocean state, we aim to isolate and quantify the influence of anthropogenic forcing on simulated climate trends. The experiments presented here focus on May start dates, which are particularly suitable for investigating boreal summer extratropical circulation trends and previously identified trend discrepancies in the tropical Pacific (Mayer et al. 2025).

These simulations represent a novel complement to traditional CMIP-style historical and scenario experiments and open new opportunities for attribution and process-based investigations of climate trends. The primary objective of this paper is to describe the experimental design and provide an initial evaluation of the resulting climate simulations. We further assess the sensitivity of SST and circulation trends in the tropical Pacific to the applied forcing perturbations. Subsequent studies (Befort et al. 2026; Maddison et al. 2026, both in preparation) will investigate the sensitivity and underlying mechanisms of boreal summer extratropical circulation trends in greater detail.

The remainder of the paper is structured as follows. Section 2 describes the experimental scenarios and model configuration. Section 3 details the atmospheric forcing perturbations. Section 4 presents the methodology used to identify the anthropogenically forced signal and to modify the ocean initial conditions accordingly. Section 5 describes the observational reference datasets and statistical methods. Results are presented in section 6, including an evaluation of the simulated climate and trend responses to the forcing perturbations. Section 7 provides a summary and conclusions.

## 2. Experimental scenarios and model configuration

We perform a suite of counterfactual hindcasts (i.e., retrospective forecasts) for the period 1993-2023 designed to simulate either amplified or muted anthropogenically forced climate change. The experiments modify the prescribed atmospheric radiative forcing ($CO_2$ and aerosols) and, in some experiments, the ocean initial conditions (ICs) as well. Adjustments to the ocean ICs are required because the ocean state at initialisation should be consistent with the cumulative effects of the imposed forcing perturbations prior to the forecast start date.

For example, in a scenario with amplified forcing, a hindcast initialised in May 2000 should begin from an ocean state that incorporates the additional heat uptake associated with the enhanced forcing between 1993 and 1999. Conversely, in a scenario with suppressed forcing, the anthropogenically forced warming accumulated over the same period should be removed from the initial ocean state. The methodology used to estimate and apply these adjustments is described in Section 4. Modifications to atmospheric initial conditions are not considered because their influence is rapidly lost during the first weeks of a seasonal forecast, whereas the ocean retains memory on substantially longer timescales.

Table 1 summarizes all experiments considered in this study. The core experiment set consists of a control simulation (CTRL), which uses observed radiative forcing and unmodified initial conditions, together with two $CO_2$-forcing counterfactuals:

- **Double $\Delta CO_2$/ocean**, in which annual $CO_2$ increments after 1993 are doubled and ocean initial conditions are adjusted accordingly;
- **Constant $CO_2$/ocean**, in which atmospheric $CO_2$ concentrations are held fixed at 1993 levels and ocean initial conditions are modified accordingly.

For these experiments, the oceanic forced signal used to perturb the initial conditions is derived from the Institute of Atmospheric Physics (IAP) ocean temperature dataset over 1975-2023 (Cheng et al. 2024, Section 4).

To isolate the direct impact of atmospheric forcing during the forecast period, we additionally perform a sensitivity experiment (**Double $\Delta CO_2$**) in which annual $CO_2$ increments are doubled after 1993 while ocean initial conditions remain unmodified. These are not to be confused with the IC perturbations related to the uncertainty representation in the ocean reanalysis (Chrust et al. 2025). This experiment quantifies the influence of enhanced $CO_2$ forcing acting only during the seven-month forecast integrations.

For the constant-forcing scenario, an additional experiment (**Constant $CO_2$/ocean/ClimLand**) employs climatological land initial conditions derived from the ERA5 climatology for 1983–2003. This simulation is used to assess the impact of land initial conditions on trends in addition to CO2 forcing ocean ICs.

The role of anthropogenic aerosols is assessed through two further experiments. In **Double ΔAER**, aerosol forcing changes after 1993 are doubled while all initial conditions remain unchanged. In **Double Δ$CO_2$/ocean/AER**, doubled aerosol changes are combined with the Double Δ$CO_2$/ocean experiment. As discussed in Section 6.4, the impact of enhanced aerosol forcing proved relatively modest; consequently, complementary experiments with fixed aerosol concentrations were not pursued.

Estimation of the forced ocean signal is subject to uncertainty. To assess the sensitivity of the results to its specification, we repeat the Double Δ$CO_2$/ocean and Constant $CO_2$/ocean experiments using an alternative estimate derived from the ORAS6 ocean reanalysis over 1993–2023 (Zuo et al. 2024). These experiments are denoted **Double Δ$CO_2$/ocean (ORAS6)** and **Constant $CO_2$/ocean (ORAS6)**, respectively.

| Experiment name | Description |
|---|---|
| Control (CTRL) | Control forecast using unperturbed radiative forcing and unperturbed ocean ICs |
| Constant $CO_2$/ocean | Removes the forced pattern of climate change from the ocean ICs after 1993 and fixes $CO_2$ concentrations at their 1993 level |
| Double Δ$CO_2$/ocean | Doubles the forced pattern of climate change in the ocean ICs after 1993 and doubles the annual rate of $CO_2$ increase after 1993 |
| Double Δ$CO_2$/ocean/AER | Doubles the forced pattern of climate change in the ocean ICs after 1993 and doubles the annual rates of both $CO_2$ increase and anthropogenic aerosol change after 1993 |
| Double Δ$CO_2$ | Doubles the annual rate of $CO_2$ increase after 1993 while leaving the ocean ICs unperturbed |
| Double $\Delta$AER | Doubles the annual rates of anthropogenic aerosol change after 1993 while leaving the ocean ICs unperturbed |
| Constant $CO_2$/ocean (ORAS6) | Sensitivity experiment equivalent to Constant $CO_2$/ocean, but using 1993–2023 ORAS6 3D ocean temperature data to estimate the forced climate-change signal. |
| Double $\Delta CO_2$/ocean (ORAS6) | Sensitivity experiment equivalent to Double $\Delta CO_2$/ocean, but using 1993–2023 ORAS6 3D ocean temperature data to estimate the forced climate-change signal. |
| Constant CO2/ocean/ClimLand | As Constant $CO_2$/ocean, but additionally replaces land ICs with a constant climatology derived from ERA5 over 1983–2003. |

**Table 1. Description of experimental scenarios**

***Model configuration***

All experiments are conducted with ECMWF's coupled forecasting system using a development version of the Integrated Forecasting System (IFS) cycle 49R2 for the atmosphere and land surface. This model configuration forms the basis of the forthcoming SEAS6 seasonal forecasting system, the successor to the currently operational SEAS5 system (Johnson et al. 2019) that uses cycle 43R1. Differences between the development version used here and the final operational SEAS6 configuration are expected to have only minor impacts on forecast characteristics.

Relative to cycle 49R1, that latest documented model cycle of the IFS (https://www.ecmwf.int/en/publications/ifs-documentation), cycle 49R2 is primarily distinguished by its coupling to version 4 of the NEMO ocean model and the SI3 sea-ice model (Madec and the NEMO System Team 2024; Ortega et al. 2025). In our experiments, the IFS is run at horizontal resolution Tco199 (corresponding to a grid spacing of around 50km at the equator), with 137 levels in the vertical. NEMO4-SI3 is run with a horizontal resolution of 0.25°, with 75 levels in the vertical.

Atmospheric initial conditions are taken from the ERA5 reanalysis. Ocean and sea-ice initial conditions are obtained from a near-final version of ECMWF's ORAS6 ocean–sea-ice reanalysis (Zuo et al. 2024); differences relative to the final release are not expected to affect the conclusions presented here.
For each experiment, 51-member ensemble hindcasts are initialized on 1 May of every year from 1993 to 2023 and integrated for 7 months lead time.

## 3. Atmospheric radiative forcing

As outlined above, we construct two counterfactual greenhouse-gas forcing scenarios. In the first scenario, atmospheric $CO_2$ concentrations are held constant at their 1993 value, representing a hypothetical world in which atmospheric CO2 concentrations remained at their 1993 level despite ongoing anthropogenic emissions in the real world. In the second scenario, the increase in atmospheric $CO_2$ is amplified by doubling the year-to-year rate of change relative to the baseline evolution after 1993.
The control $CO_2$ concentrations used in the IFS follow the CMIP6 historical forcing dataset up to 2014 and the moderate SSP3-7.0 scenario thereafter. Figure 2 shows the unperturbed $CO_2$ evolution used in the model through 2040, together with the two counterfactual scenarios described above. For reference, the figure also includes the CMIP6 SSP2-4.5 (moderate) and SSP5-8.5 (extreme) scenarios. By 2023, the amplified $CO_2$ scenario reaches concentrations comparable to those projected under SSP5-8.5 in the late 2030s, illustrating the magnitude of the imposed perturbation.

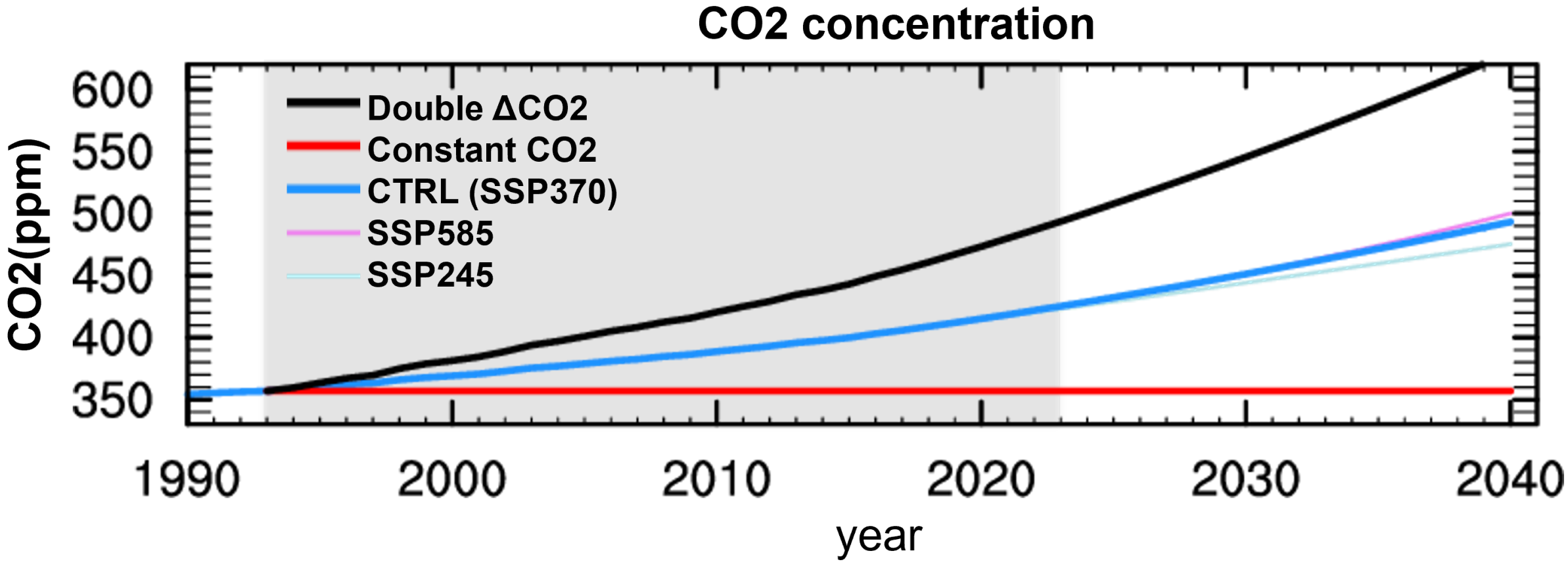

**Figure 2. Comparison of historical and scenario-type atmospheric CO2 concentration evolution with the prescribed evolutions in the counterfactual hindcasts. The grey shading indicates the hindcast period 1993-2023.**

For tropospheric aerosols, the baseline forcing is derived from a time-varying climatology based on emission data from the Community Emissions Data System (CEDS; Stockdale et al. 2022). Because the CEDS emissions dataset available during the production of the forcing fields extended only to 2019, the aerosol forcing used in the model becomes effectively constant after approximately 2015 as a result of the processing methodology used to generate the climatology.

To investigate the impact of enhanced aerosol forcing, we construct a perturbed aerosol dataset using an approach analogous to that employed for the amplified $CO_2$ scenario. However, because the temporal evolution of many aerosol species is non-monotonic, simply doubling the year-to-year changes can produce negative concentrations in some regions and periods. To avoid such unphysical situations, the perturbed aerosol dataset is generated in two steps. First, the temporal rates of change of all aerosol species are doubled throughout the record. Second, any negative concentrations introduced by this procedure are reset to zero. Figure 3 illustrates the resulting evolution for four representative aerosol species.

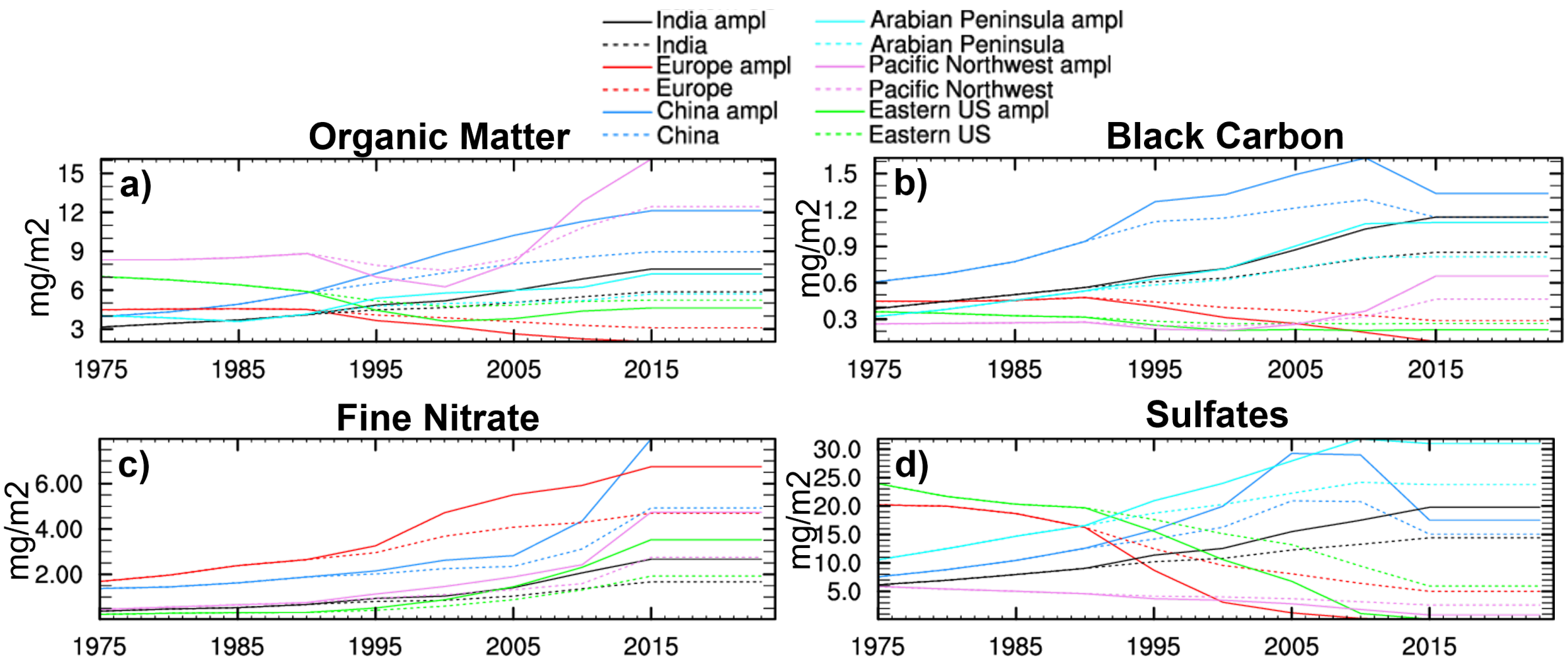

**Figure 3. Unperturbed (as used in runs without aerosol perturbations) and perturbed evolution (as used in ∆AER and ∆CO2/ocean/AER) for four different aerosol species in different regions.**

## 4. Determining forced patterns and preparation of initial conditions

In free-running coupled simulations, Earth system components continuously respond to changes in external forcing. One important consequence is the accumulation of excess heat in the ocean under increasing greenhouse-gas concentrations (e.g., Von Schuckmann et al. 2023). To ensure consistency between the imposed forcing perturbations and the initialized state of the climate system, the ocean and sea-ice ICs used in the counterfactual hindcasts must therefore be adjusted. For example, in the amplified-$CO_2$ scenario, ocean temperatures at a given start date must be increased to reflect the accumulated additional warming (heat uptake) resulting from the enhanced anthropogenic forcing up to that time.

Anthropogenically forced climate patterns can be estimated either from coupled climate-model simulations with perturbed anthropogenic forcings (e.g., Fasullo et al. 2024) or statistically from the historical record (e.g., Wills et al. 2026). The model-based approach has the advantage of providing dynamically consistent forced responses and, when derived from the same model used for the hindcasts, reflects the model's own sensitivity to external forcing. However, such estimates are affected by biases in the simulated forced response. A prominent example is the widespread inability of coupled climate models to reproduce the observed trends in the tropical Pacific, with a strengthening of the zonal SST gradient and associated strengthening easterly winds (e.g., Seager et al. 2022; Rugenstein et al. 2023; Mayer et al. 2025). Because our hindcasts are initialized from observationally constrained states, we adopt an observationally based estimate of the forced climate signal.

A limitation of statistical approaches is the relatively short and spatially sparse observational record, particularly for subsurface ocean temperatures. Although several gridded three-dimensional ocean temperature datasets extend back more than a century, including Hadley EN4 (Good et al. 2013) and IAP (Cheng et al., 2024), uncertainties increase prior to the satellite era (Cheng et al., 2024). Estimating forced patterns therefore requires a compromise between record length, which increases statistical robustness, and data quality and temporal homogeneity, which are essential for producing consistent initial conditions and reliable trend estimates.

The forced pattern in 3D ocean temperature is estimated following Leach et al. (2024), with several modifications. We first compute the Anthropogenic warming index (AWI; Haustein et al. 2017) from the regression of anthropogenic radiative forcing [estimated using the Finite Amplitude Impulse Response simple climate model (FaIR) in version 2.0; Leach et al. 2021a] with observed global near-surface temperature changes based on HadCRUT5 (Morice et al. 2021) over the period 1850-2022. The AWI provides an estimate of global temperature changes attributable to anthropogenic changes in atmospheric composition. The anthropogenically forced pattern in 3D ocean temperatures is then obtained by regressing 3D ocean temperature observations against the AWI.

We tested multiple observational datasets and regression periods and found that the estimated forced patterns converge when data from approximately the mid-1970s onward are used. We therefore derive the final 3D pattern from the IAP ocean temperature dataset over 1975–2023. Unlike Leach et al. (2024), who combined independently estimated surface and subsurface patterns, we derive a single three-dimensional pattern from a common dataset and period, thereby avoiding potential spatial inconsistencies between surface and subsurface responses. The resulting pattern can be interpreted as the 3D ocean temperature increment associated with a unit change in the AWI.
The forced pattern represents the integrated response to all anthropogenic forcings, whereas our experiments modify only $CO_2$ and aerosol forcing. This introduces a degree of inconsistency, as changes in other forcings, such as methane, are not explicitly represented. However, we expect the resulting error to be small relative to the magnitude of the imposed perturbations and the timescales considered here.

For each hindcast start date, perturbed ocean ICs are generated by adding or subtracting the forced pattern scaled by the AWI change since 1993. Consequently, the imposed perturbation is smallest for forecasts initialized shortly after 1993 and increases progressively towards 2023. To preserve pressure gradients and thus balanced ocean currents in the unperturbed ocean initial conditions, the 3D salinity field is adjusted such that densities remain unchanged, as in Leach et al. (2024).

The anthropogenically forced sea-ice response is estimated using an analogous methodology. Forced patterns are estimated for sea-ice concentration (SIC) and sea-ice thickness (SIT). Owing to the large uncertainties in historical sea-ice thickness observations, we use ORAS6 data covering 1993-2023 to determine the forced patterns in SIC and SIT. The resulting SIC and SIT increments are applied to the initial conditions in the same manner as for ocean temperature. The increments are distributed across the five sea-ice categories following the approach used for assimilation increments (Browne et al. 2026), and extensive quantities such as ice enthalpy are scaled to preserve the associated intensive properties.

Atmospheric initial conditions are not modified. Constructing dynamically and thermodynamically consistent atmospheric perturbations would be considerably more complex than for the ocean and sea ice. Furthermore, because of the comparatively small heat capacity and short memory of the atmosphere, we expect atmospheric temperature and humidity fields to adjust rapidly to the perturbed lower boundary conditions. Consistent with Leach et al. (2021b), we assume that thermodynamic equilibrium between the atmosphere, ocean, and sea ice is largely re-established within the first few weeks of the hindcast integrations.

Figure 4a shows estimates of the forced pattern used to perturb the ICs in the main experiments. At the surface, the pattern exhibits widespread warming, with notable exceptions. The central equatorial and southeastern Pacific show a neutral forced pattern, and the Southern Ocean displays a negative forced SST response. The zonal mean in Fig. 4a shows that the maximum forced SST warming in May is found in the northern extratropics, while the zonal mean forced signal in the Southern Ocean is negative. In the subsurface tropical Pacific, the estimated forced pattern features pronounced warming in the western Pacific, consistent with a strengthening zonal temperature gradient and enhanced easterly trade winds.

To assess the robustness of the estimated forced pattern, we compare the SST response with the forced SST patterns estimated by Wills et al. (2026), who tested the capabilities of a wide range of statistical methods to detect the forced signal. Figure 4c shows the average SST pattern from the five best-performing statistical methods identified in that study. The large-scale patterns are generally in good agreement, and our pattern lies within the range of the five best-performing methods. The mean spatial correlation between our SST pattern and the 27 methods classified as skillful by Wills et al. (2026) is 0.74, increasing to 0.80 when only the five best-performing methods are considered. This provides confidence that the methodology employed here yields reasonable estimates of the forced SST response.

We further note that the estimated forced SST pattern closely resembles the simple linear SST trend over 1975–2023 (Figure S1). This suggests that the regression-based method does not attribute a substantial fraction of the observed trend to slowly varying modes of internal variability. Indeed, as illustrated by Wills et al. (2026), the different methods exhibit a wide range in the fraction of internal variability contributing to the observed SST trends. This large range can be interpreted as the uncertainty inherent to the statistical methods to determine forced signals.

Figure S1 also compared forced SST patterns derived using data covering the shorter 1993-2023 period. Both the IAP-based forced SST pattern as well as the IAP-based SST trend are stronger compared to the estimates based on 1975-2023. This suggests that using the 1975-2023 forced ocean patterns may underestimate the forced trend during the 1993-2023 period. Figure S1 also compares the forced SST pattern based on IAP and ORAS6 data 1993-2023: these are very similar.

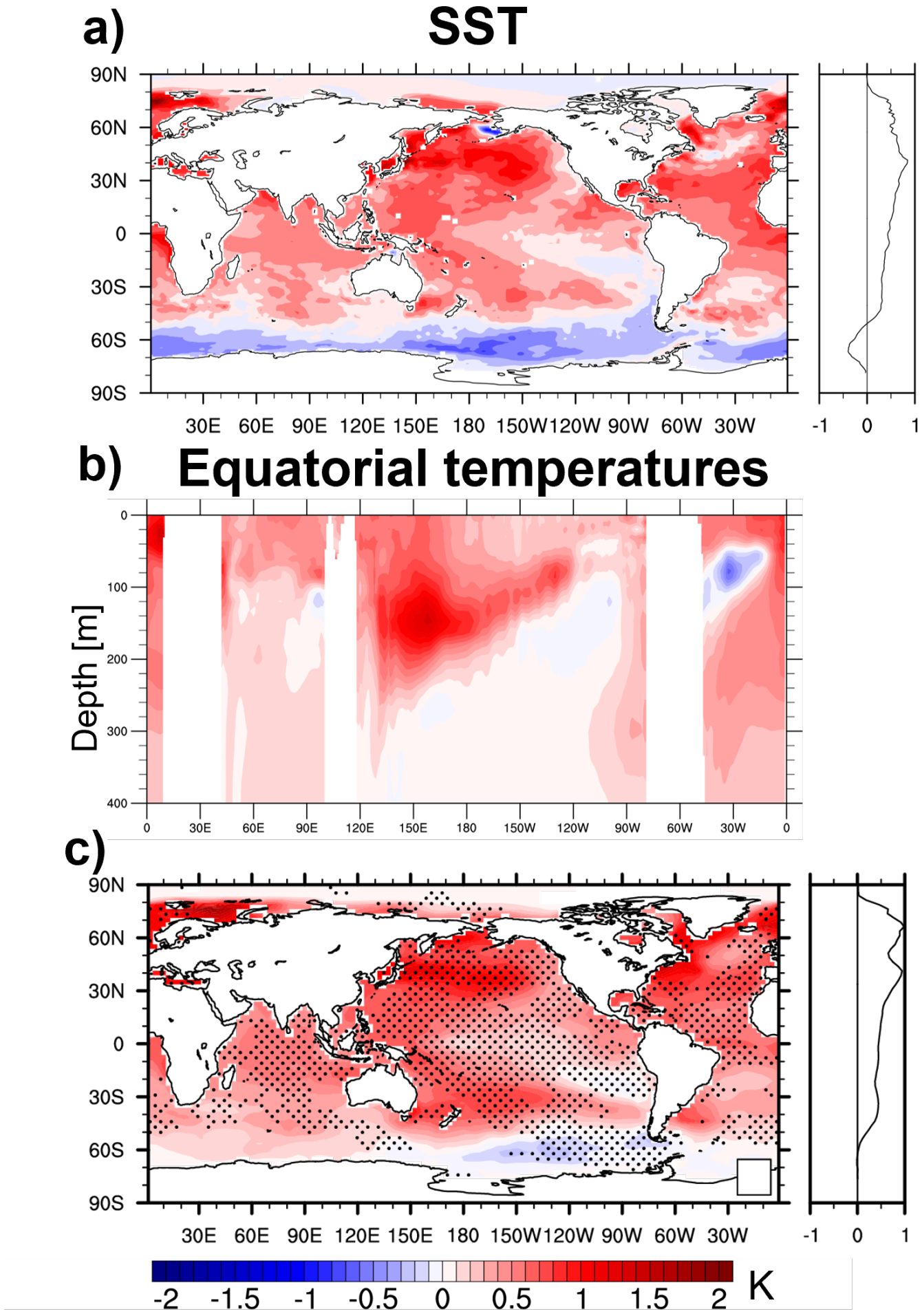

**Figure 4. Anthropogenically forced pattern estimate in (a) SST and (b) subsurface temperature 2023/05 relative to 1993/05 (based on IAP data 1975-2023); (c) as (a), but averaged over 5 forced pattern estimates using the best 5 best-performing methods according to Wills et al (2026); stippling in (c) shows where our estimate is within the range of the 5 best estimates used for the pattern shown in (c).**

## 5. Validation data and statistical methods

Atmospheric variables, including winds and temperatures as well as SSTs, are validated against ERA5 reanalysis (Hersbach et al. 2020). The Niño3.4 index is defined as the area-averaged SSTs over 5°S-5°N, 170°W-120°W. Top-of-atmosphere radiative fluxes are evaluated using observations from the Clouds and the Earth's Radiant Energy System Energy Balanced and Filled dataset (CERES-EBAF; Loeb et al. 2018), available from 2001 onward. Precipitation is validated against version 3.2 of the Global Precipitation Climatology Project dataset (GPCP; Huffman et al. 2023). Ocean heat content (OHC) is based on the preliminary ORAS6 ocean reanalysis (Zuo et al. 2024), which also provides the ocean initial conditions used in the hindcasts.

Linear trends are estimated using ordinary least-squares regression. Trend uncertainties are calculated from the standard error of the regression slope while accounting for temporal autocorrelation. Unless stated otherwise, uncertainties are reported as 90% confidence intervals.

To assess the consistency between observed trends (a single realisation) and the range of trends simulated by the hindcasts (an ensemble of many plausible trends), we follow the methodology of Mayer et al. (2025). We generate 1,000 synthetic hindcast time series by randomly selecting one ensemble member for each forecast year. This assumes that each ensemble member is an equally plausible outcome of the seasonal climate evolution and that the resulting temporal sequence of seasonal anomalies provides time series with realistic characteristics.

A linear trend is calculated for each of the 1,000 synthetic time series, yielding a distribution of plausible hindcast trends. Observed and simulated trends are considered statistically consistent if the observed trend lies within the central 90% of the hindcast trend distribution.

# 6. Evaluation

## 6.1. Global mean quantities

We begin our assessment of the counterfactual hindcast by examining the evolution of global-mean 2m temperature (T2M), SST, and ocean heat content (OHC).
Figure 5a shows global-mean T2M evolution in JJA relative to JJA 1993. The CTRL simulation closely reproduces the observed evolution from ERA5, both in terms of long-term trends (0.23 ± 0.03 K/dec and 0.23 ± 0.05 K/dec, respectively) and interannual variability. The temporal correlation between detrended ensemble mean and ERA5 is r=0.79 (r=0.96 when trends are not removed). In particular, the exceptional warmth of 2023 is captured well, likely because part of the signal was already present in the initial conditions on 1st May 2023.

The Double $\Delta CO_2$/ocean experiment exhibits a substantially larger warming trend (0.37 ± 0.03 K/dec) while retaining nearly unchanged interannual variability. The detrended correlation with ERA5 is r=0.80, and the correlation with CTRL is r=0.99. This indicates that amplifying the forced signal in the ocean ICs primarily shifts global temperatures to a warmer baseline without degrading forecast skill for year-to-year anomalies.

The Constant CO2/ocean exhibits much weaker T2m warming trends (0.09 ± 0.03 K/dec), although temperatures continue to increase over the hindcast period. The detrended correlation with CTRL remains high (r=0.80), indicating that much of the interannual variability is preserved. One possible explanation for the residual warming trend is the use of unmodified atmospheric and land initial conditions, which contain an anthropogenic warming signal that may influence temperatures during the forecasts. However, an additional experiment using climatological land initial conditions (Constant $CO_2$/ocean/ClimLand) produces only a slightly weaker trend (0.08 ± 0.03 K/dec; not shown). While land initial conditions contribute to trends over northern extratropical land and, to a lesser extent, oceanic regions (Figure S2), they cannot explain all of the remaining global warming. A small contribution may also arise from the potential underestimation of the forced signal during the 1993-2023 period by our estimate (based on 1975-2023, see figure S1 and discussion in section 4). However, the global JJA T2M trend in Constant $CO_2$/ocean_ORAS6 still is 0.07 ± 0.03 K/dec, i.e. only slightly reduced compared to Constant $CO_2$/ocean. Another likely contributor is a remaining warming signal in the subsurface ocean ICs, see Figs 5c and d discussed further below.
The Double $\Delta CO_2$ experiment, in which only atmospheric $CO_2$ concentrations are modified, produces a T2M trend of 0.26 ± 0.03 K/dec, very similar to CTRL. This confirms that the direct impact of enhanced radiative forcing on seasonal timescales is small and that most of the response arises from the altered ocean ICs, which embody the accumulated effects of past forcing changes.
The evolution of global-mean SSTs (Fig. 5b) closely mirrors that of T2M. ERA5 SSTs warm at 0.17 ± 0.03 K/dec, and CTRL exhibits the same trend. Trends increase to 0.28 ± 0.03 K/dec in Double $\Delta CO_2$/ocean, decrease to 0.05 ± 0.03 K/dec in Constant CO2/ocean, and remain largely unchanged at 0.18 ± 0.03 K/dec in Double ∆CO2/ocean, indicating a response broadly consistent with that found for T2M.

Both upper-ocean (0–300 m) OHC (Fig. 5c) and OHC below 300 m (Fig. 5d) exhibit a more continuous increase than the surface variables. CTRL agrees well with the observationally constrained ORAS6 estimate, although it slightly underestimates the recent rate of heat uptake according to ORAS6 (0.17 ± 0.01 × $10^8$ J/m²/dec). Similar to the surface variables, 0-300m OHC trends increase substantially in Double $\Delta CO_2$/ocean (0.25 ± 0.01 × $10^8$ J/m²/dec) and are strongly reduced in Constant $CO_2$/ocean (0.06 ± 0.01 × $10^8$ J/m²/dec).

A notable feature of Constant $CO_2$/ocean is the pronounced increase in OHC below 300 m between approximately 2002 and 2006. This period coincides with the rapid expansion of the Argo observing network and the resulting improvement in subsurface temperature coverage. The warming therefore likely reflects, at least in part, inhomogeneities in the underlying ocean reanalysis associated with changes in the observing system. Because this signal is distinct from the long-term average OHC increase (i.e. unrelated to the increase in radiative forcing and consequently AWI), it is not removed by our methodology for estimating and subtracting the forced ocean-temperature response.

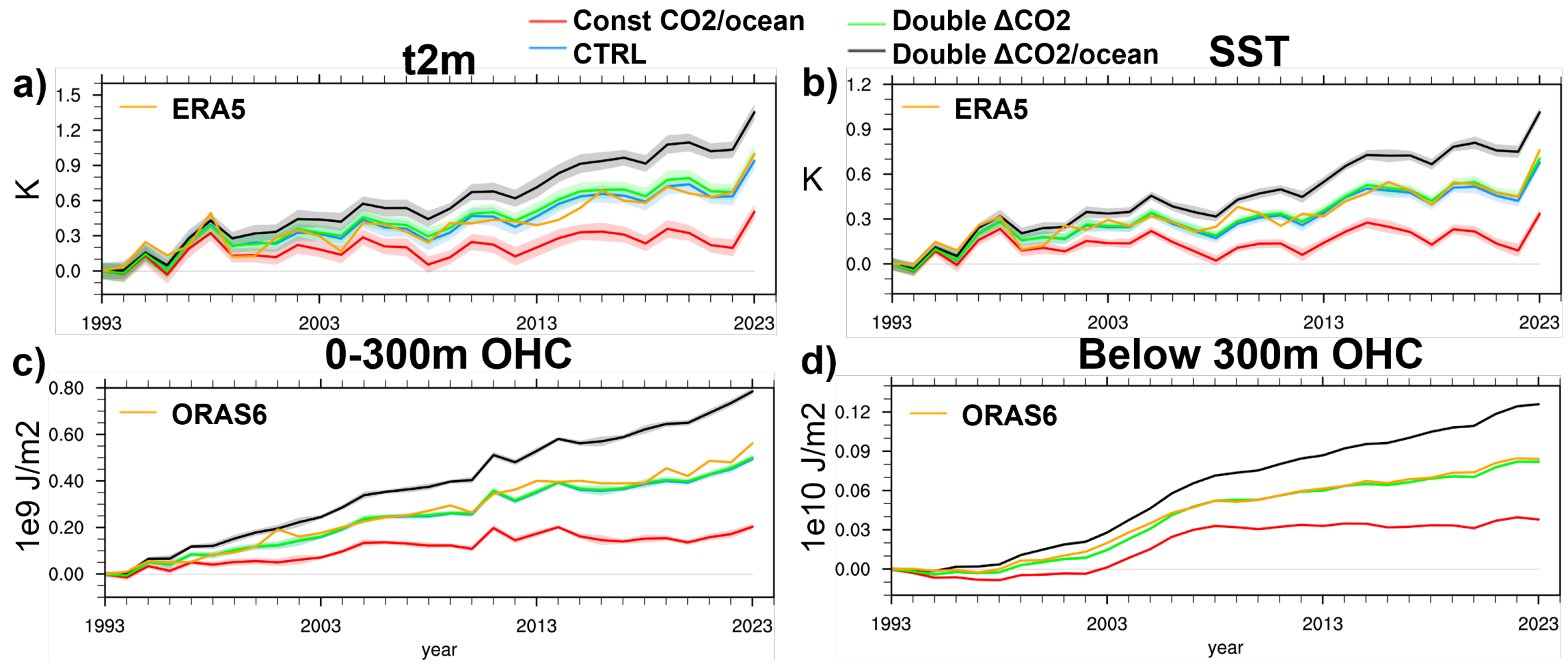

**Figure 5. Global mean evolution of JJA (a) T2M, (b) SST, (c) 0-300m OHC, and (d) full-depth OHC relative to JJA 1993 from the CTRL, a selection of counterfactual hindcasts, and ERA5 (T2M, SST) and ORAS6 (OHC).**

### 6.2. Climate mean state

Although the perturbations in our experiments grow with time, their climatic impacts can be assessed by comparing fields averaged over the final decade of the hindcasts (2014–2023), when the imposed perturbations are largest.

Figures 6a and 6b show JJA mean SST differences relative to CTRL in the Constant $CO_2$/ocean and Double $\Delta CO_2$/ocean experiments, respectively. The close resemblance between these patterns and the imposed SST perturbations (Fig. 4a) demonstrates that the model largely preserves the large-scale forced signal throughout the first few months of the forecasts. Figure 6c shows the asymmetry between positive and negative forcing perturbations, defined as the differences (Double $\Delta CO_2$/ocean − CTRL) and (CTRL − Constant $CO_2$/ocean). Most of the asymmetry is confined to the Arctic, where positive ice-albedo feedbacks are stronger in Double $\Delta CO_2$/ocean compared to Constant $CO_2$/ocean. In the positive perturbation case, sea ice is completely removed in many locations, substantially reducing surface albedo and amplifying warming. In contrast, the negative perturbation primarily increases ice thickness without greatly altering ice extent or albedo, resulting in a weaker response.
The mean response of 10-m zonal winds (u10m; Figs. 6d,e) is consistent with changes in tropical SST gradients. In Double $\Delta CO_2$/ocean, the strengthened east–west SST gradient across the tropical Pacific is accompanied by stronger easterly trade winds over the central equatorial Pacific. Conversely, the weakened SST gradient in Constant $CO_2$/ocean is associated with weaker easterlies. The asymmetry diagnostic (Fig. 6f) indicates that in the equatorial Pacific the wind response to the negative forcing perturbation is stronger than the response to the positive perturbation.

A similar asymmetry is evident in tropical precipitation. Over the Indo-Pacific warm pool, the precipitation reduction in Constant $CO_2$/ocean exceeds the precipitation increase in Double $\Delta CO_2$/ocean. This behavior is consistent with energetic constraints on tropical precipitation, which limit the increase in rainfall that can accompany SST warming, whereas precipitation reductions associated with cooling are less strongly constrained (Muller and O'Gorman 2011).

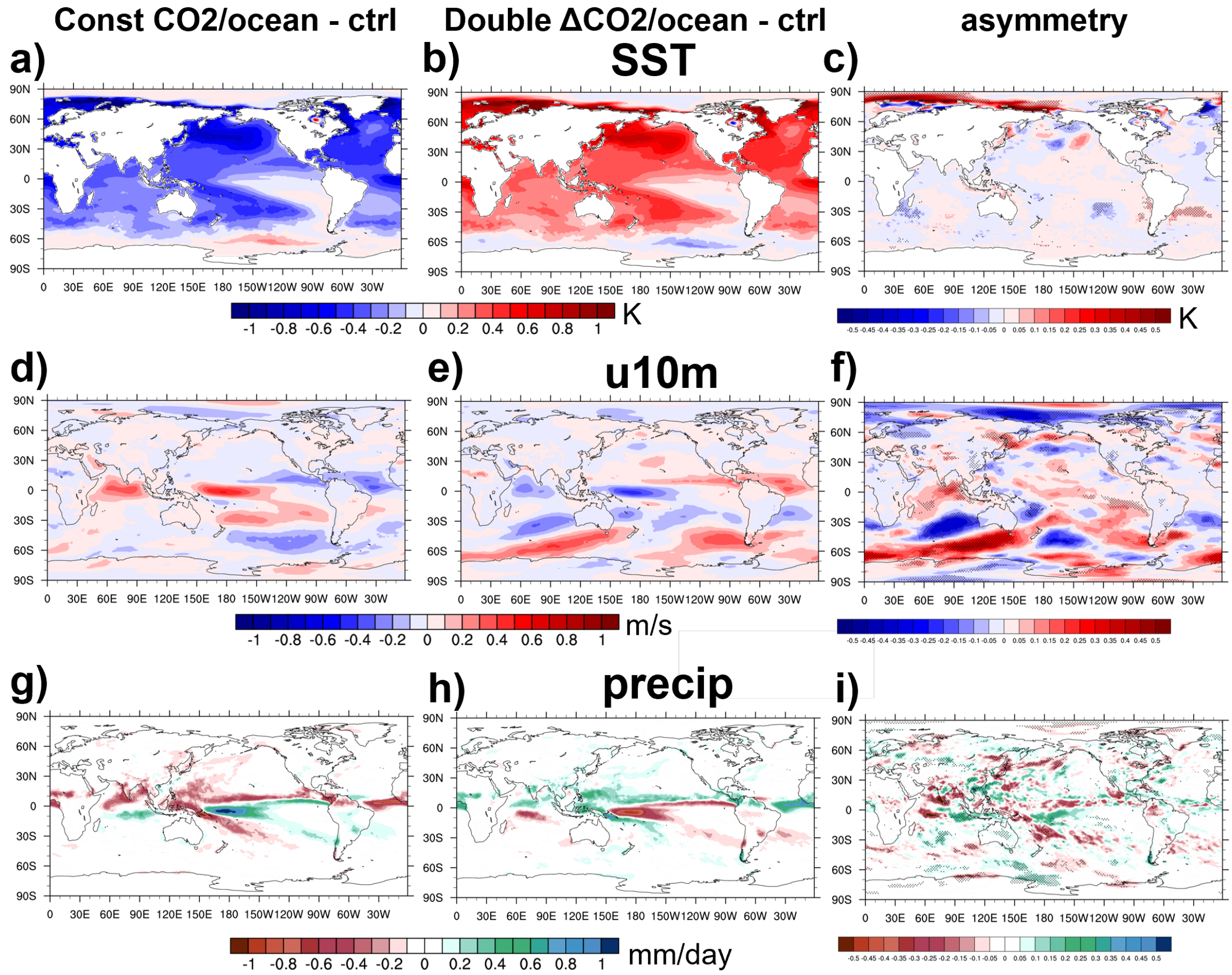

**Figure 6. JJA 2014-2023 mean ensemble mean difference between (left column) Constant CO2/ocean and CTRL, (middle column) Double ΔCO2/ocean and CTRL, and (right column) the asymmetry [defined as the differences (Double Δ$CO_2$/ocean − CTRL) and (CTRL − Constant $CO_2$/ocean)] of the differences. The first row shows differences in SST, second row differences in u10m, and third row differences in precipitation.**

### 6.3. SST seasonality

We next examine how forcing and IC perturbations affect the SST drift during the final decade of the hindcasts. Figure 7 summarizes changes in the seasonal SST cycle for two domains: the global ocean and the Niño 3.4 region.

Globally, the CTRL simulation exhibits too strong seasonal SST warming (by ~0.04K until August) and too weak seasonal SST cooling relative to observations, yielding a drift bias by ~0.08K from May to November (Fig. 7a). However, our primary interest is the change in model drift relative to CTRL. The experiments with perturbed ocean ICs display seasonal

SST evolution that is remarkably similar to CTRL (change in May to November drift bias ~0.01K), indicating that the imposed ocean perturbations do not substantially alter the model's mean drift characteristics (albeit starting from a different level due to the ocean perturbations). In contrast, Double $\Delta CO_2$, in which only the atmospheric $CO_2$ forcing is modified, exhibits a substantially stronger positive SST drift (~0.04K change in May to November drift bias compared to CTRL). This suggests that the combined perturbation of forcing and ocean initial conditions yields a more balanced model response than modifying the radiative forcing alone.

The drift changes exhibit substantial spatial structure that is physically consistent with the imposed perturbations and is discussed in detail in Figure S3 and associated text. Here, we focus on the Niño-3.4 region, where atmosphere–ocean coupling is particularly strong (Fig. 7b).

In the Niño-3.4 region, CTRL underestimates the observed seasonal cooling. Nevertheless, the experiments with adjusted ocean initial conditions show only minor deviations from CTRL throughout the forecast period, including at the longest lead times. This indicates that the IC perturbations do not introduce substantial mean-state biases or drift changes in the tropical Pacific, despite the strong local atmosphere–ocean coupling. The persistence of similar drift characteristics across the experiments provides additional evidence that the imposed ocean perturbations are dynamically consistent with the model climatology.

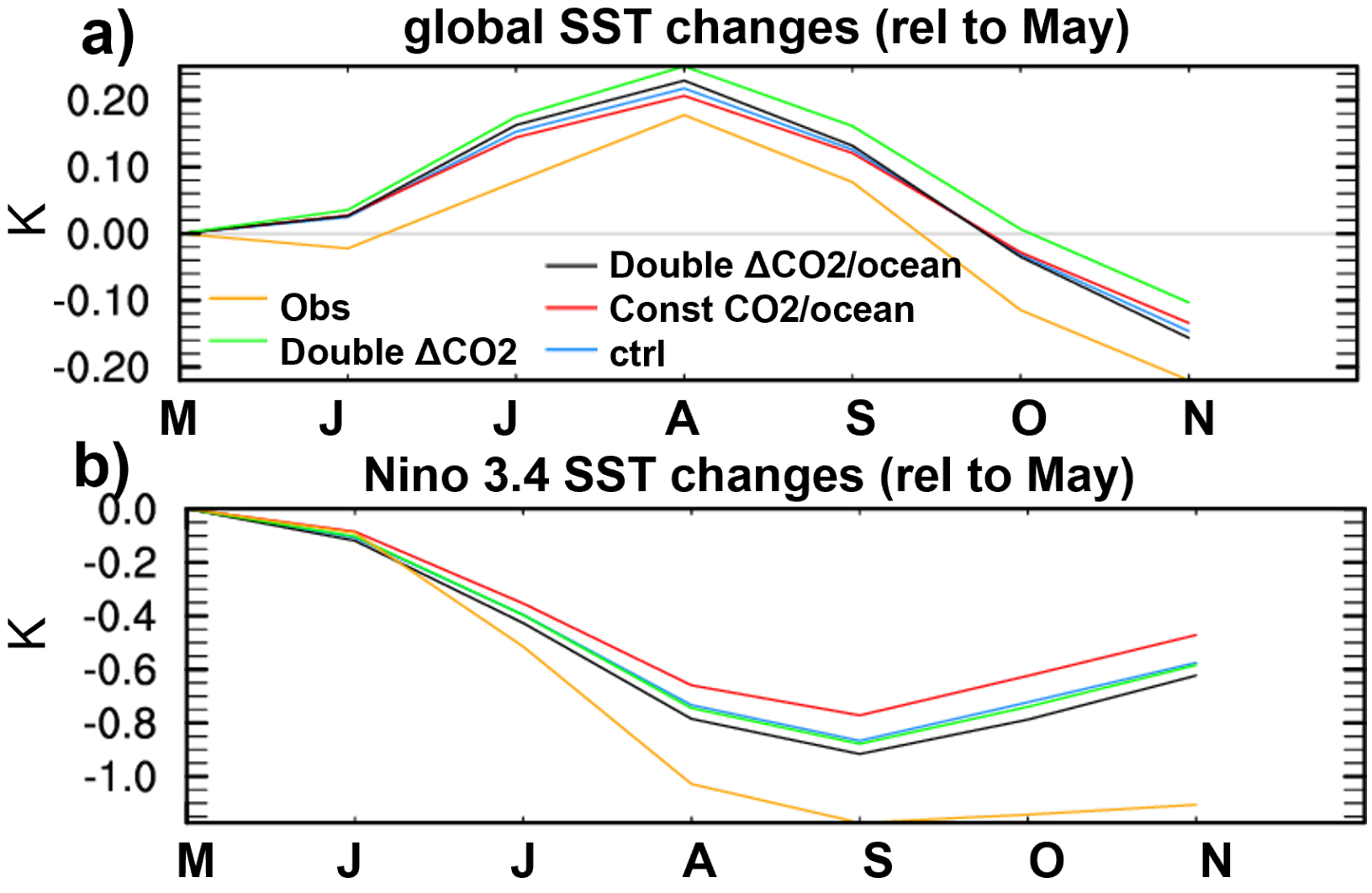

**Fig. 7. Area-averaged monthly SST changes relative to May (2014-2023): (a) global mean, (b) Nino 3.4 region.**

Fig. 8 shows that the counterfactual forecasts exhibit comparable predictive skill for SST anomalies in the Nino 3.4 region to the control, both in terms of RMSE and correlation skill. The slightly reduced RMSE of Constant CO2/ocean compared to the CTRL arises from the fact that the CTRL exhibits a spurious warming trend in the Nino 3.4 region similar to the operational SEAS5 (Mayer et al. 2025), which is ameliorated in Constant CO2/ocean (see also section 6.5 below). The good ENSO skill in the counterfactual hindcasts indicates that the ocean perturbations do not have detrimental effects on zonal gradients in the tropical Pacific. Thus, the counterfactual hindcasts are very well-suited to investigate the evolution of specific ENSO events under different climate conditions which will be the focus of future work.

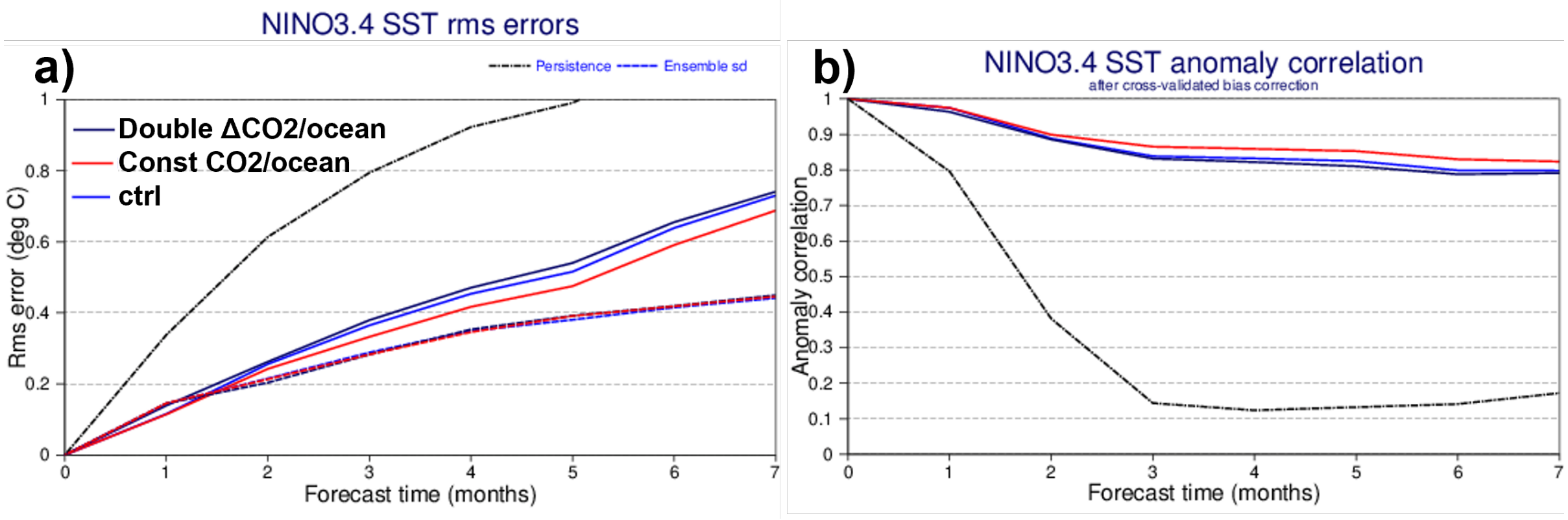

**Fig. 8. Prediction skill of Nino 3.4: (a) bias-corrected RMSE and (b) anomaly correlation skill.**

### 6.4. Top-of-atmosphere fluxes

We next investigate how the forcing and IC perturbations affect top-of-atmosphere (TOA) radiative fluxes. Figure 9 shows the evolution of global-mean JJA net TOA flux, absorbed shortwave radiation (ASR), and outgoing longwave radiation (defined positive downward and therefore denoted −OLR) for the CTRL and the main sensitivity experiments from 2001 onward, when satellite observations become available. Fig. 10 and Table S1 complement Fig. 9 by summarizing the corresponding global-mean TOA flux trends.

The CTRL hindcast exhibits a small negative net TOA flux bias in the early 2000s that increases with time, reaching approximately 1 W/m² in recent years (Fig. 9a). This growing bias arises because the simulated positive net TOA trend is substantially weaker than observed (0.25 ± 0.21 W/m²/dec versus 0.73 ± 0.23 W/m²/dec; Fig. 10 and Table S1).

Consistent with Loeb et al. (2024), changes in ASR are the dominant contributor to the observed TOA trend.

The CTRL simulation exhibits a negative ASR bias, indicating insufficient absorption of incoming shortwave radiation (Fig. 9b). This bias is widespread across the oceans, except in major upwelling regions and western boundary current systems (Figure S4). In addition, the CTRL underestimates the observed ASR trend (0.35 ± 0.24 W/m²/dec compared with 0.86 ± 0.23 W/m²/dec in observations). In return, the CTRL simulation has a positive -OLR bias, implying insufficient outgoing longwave radiation, and also an underestimation of the magnitude of the observed negative −OLR trend (Fig. 9c). The positive −OLR bias is particularly pronounced in the Intertropical Convergence Zone (ITCZ), where it suggests cloud tops that are too cold or too high, but is also evident in the extratropics (Figure S4). Importantly, these TOA flux biases exhibit only weak lead-time dependence (Figure S5).

The counterfactual hindcasts produce markedly different TOA trends. Relative to CTRL, Double $\Delta CO_2$/ocean increases the positive net TOA trend by approximately 40%, whereas Constant $CO_2$/ocean reduces it by a similar amount (Fig. 10 and Table S1). These changes are primarily driven by corresponding increases and decreases in ASR trends.

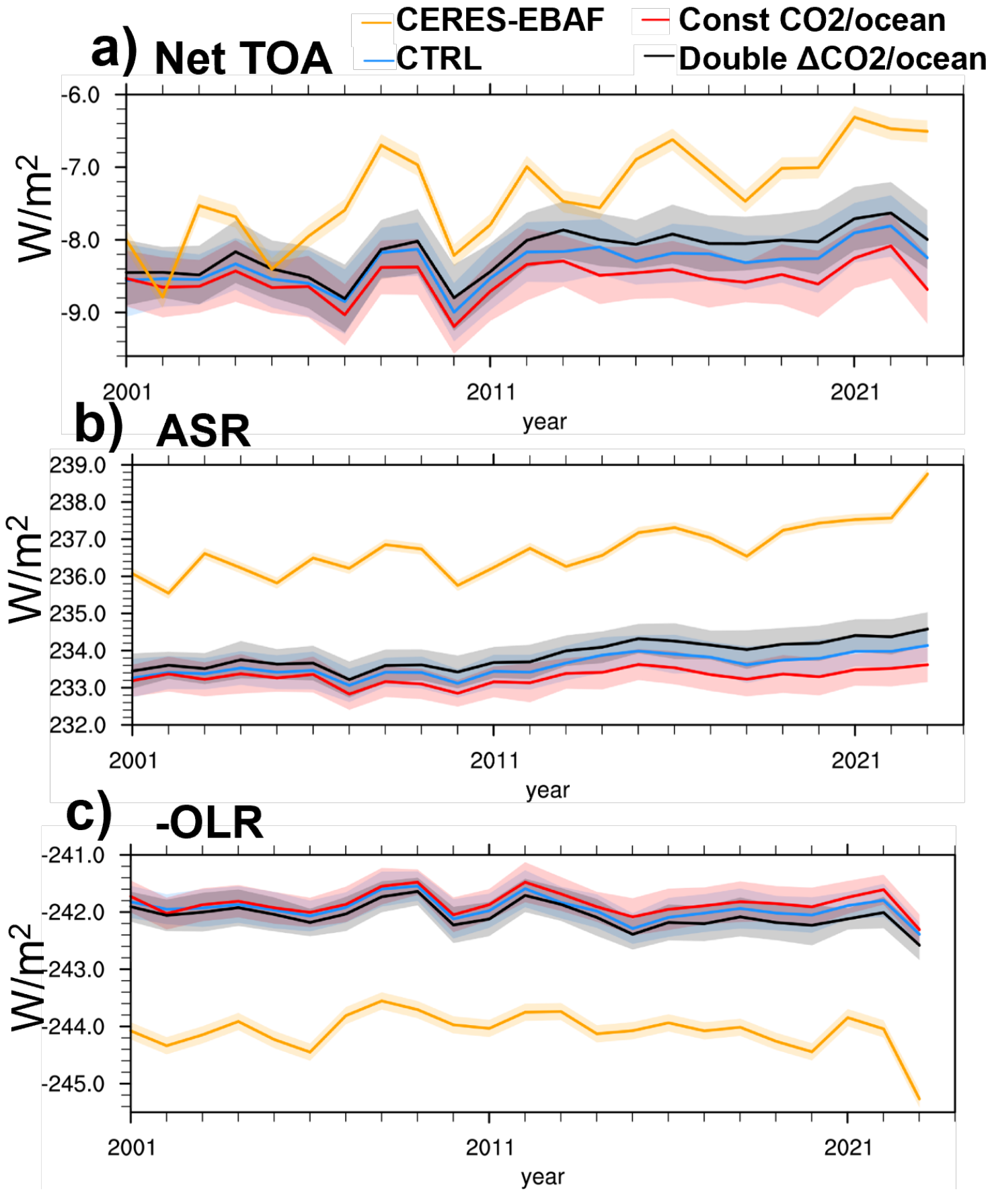

**Fig. 9. Global mean top-of-atmosphere radiation evolution in main experiments for JJA 2001-2023: (a) net radiation, (b) ASR, (c) -OLR. Trends for other experiments are provided in Table S1.**

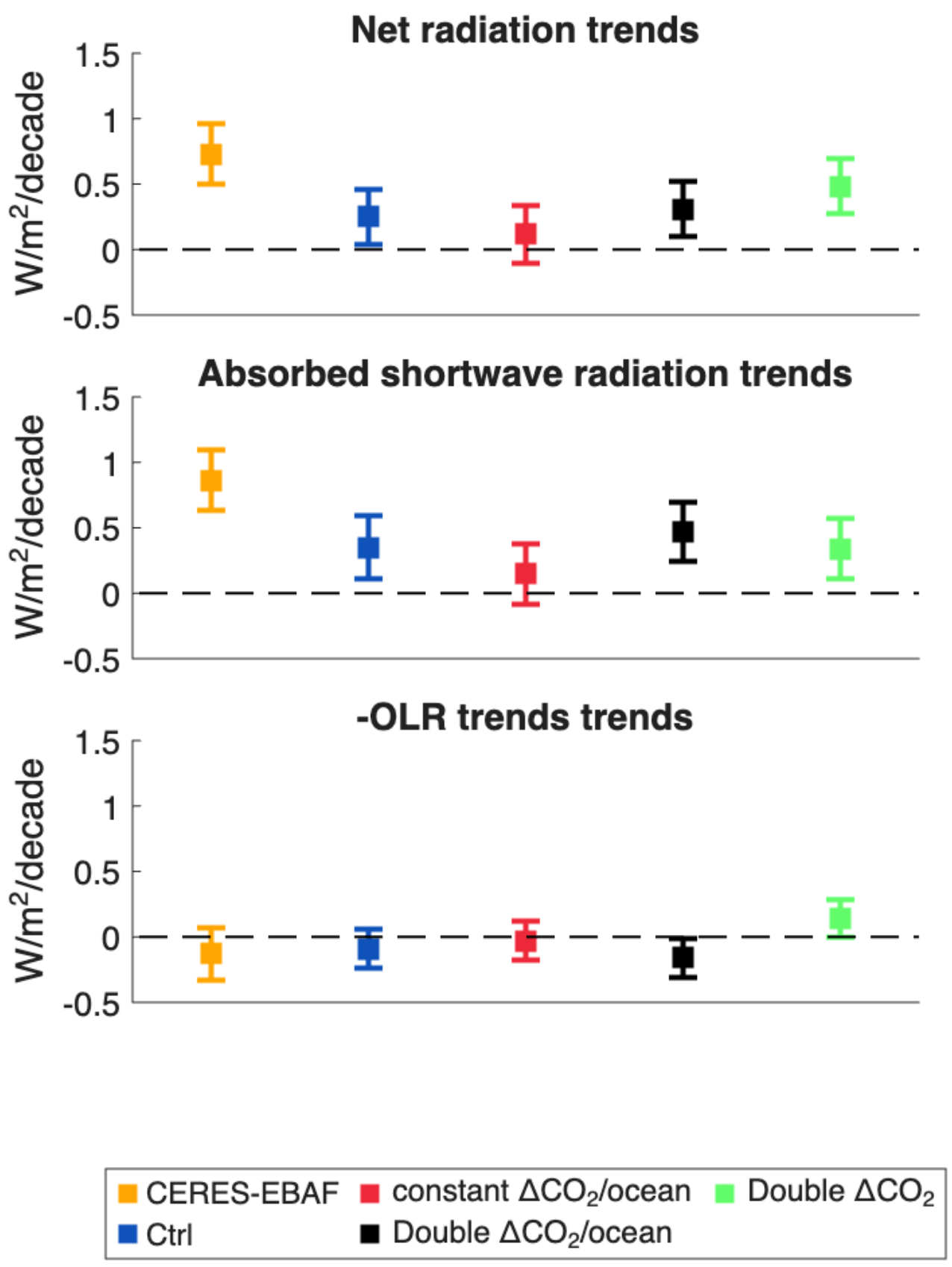

**Fig. 10. Global mean top-of-atmosphere radiative trends JJA 2001-2023 [W/m2/decade] in observations and hindcast datasets. Hindcast trend central values are ensemble mean trends, and the 90% confidence intervals (indicated by the whiskers) are obtained from the bootstrapped trend distributions. Trends for other experiments are presented in Table S1.**

In the experiments that combine modified $CO_2$ forcing with adjusted ocean ICs, the impact on global OLR trends is comparatively small. In contrast, the Double $\Delta CO_2$ experiment, which modifies atmospheric $CO_2$ concentrations alone, exhibits a substantial increase in −OLR (i.e., reduced outgoing longwave radiation). In this case, the enhanced greenhouse effect dominates the radiative response, increasing net TOA flux primarily through longwave trapping. Consequently, Double $\Delta CO_2$ achieves a more realistic net TOA trend for the wrong physical reason, highlighting the importance of adjusting ocean initial conditions alongside the radiative forcing.

To better understand these changes, Figure 11 shows maps of JJA TOA flux trends in observations, CTRL, and selected counterfactual experiments. Observed net TOA trends are predominantly positive over the Northern Hemisphere, with ASR trends accounting for most of the spatial structure (Loeb et al. 2024). The CTRL hindcast reproduces the broad pattern but systematically underestimates positive ASR and net TOA trends in key regions. Across much of the Pacific and high northern latitudes, the observed trends lie outside the ensemble trend distribution, indicating statistically significant discrepancies (see Section 5 for the approach to generate the trend distribution).

The observed −OLR trends indicate enhancement of deep convection over the Maritime Continent and reduced convection over the central equatorial Pacific (Fig. 11c), consistent with the observed strengthening of the Pacific Walker cell. In contrast, the CTRL hindcast exhibits the opposite tendency (Fig. 11f), implying a spurious weakening of the Walker circulation. This behaviour is consistent with previous findings for SEAS5 (Mayer et al., 2025).

Double $\Delta CO_2$/ocean substantially improves ASR and net TOA trends in several key regions (Figs. 11g–i), including the Pacific ITCZ and Maritime Continent, the northern extratropical Pacific, and the Arctic. In the tropical Pacific, these improvements are associated with more realistic trends in zonal winds and precipitation (Section 6.5). In the North Pacific, the improvements likely reflect interactions between cloud biases and SST trend patterns. Previous studies have shown that the IFS model underestimates low clouds and consequently overestimates ASR in this region (Forbes et al. 2016; Li et al. 2021; see also Figure S4). The low baseline cloud amount in the CTRL hindcasts may reduce cloud sensitivity to surface warming, causing CTRL to underestimate observed ASR trends. The enhanced warming imposed in Double $\Delta CO_2$/ocean (Fig. 4a) likely enhances the cloud response and thereby improves the ASR trend, although the underlying cloud bias limits the magnitude of the improvement. Consistent with this interpretation, Figure S6 shows only modest improvements in the shortwave and longwave components in the North Pacific. Improvements in the Arctic are likely linked to the imposed sea-ice perturbations. By enhancing sea-ice loss relative to CTRL, Double $\Delta CO_2$/ocean strengthens the ice–albedo feedback and thereby increases positive ASR trends. This is consistent with the known tendency of SEAS5 and related systems to underestimate long-term Arctic sea-ice decline (e.g., Batté et al. 2020).

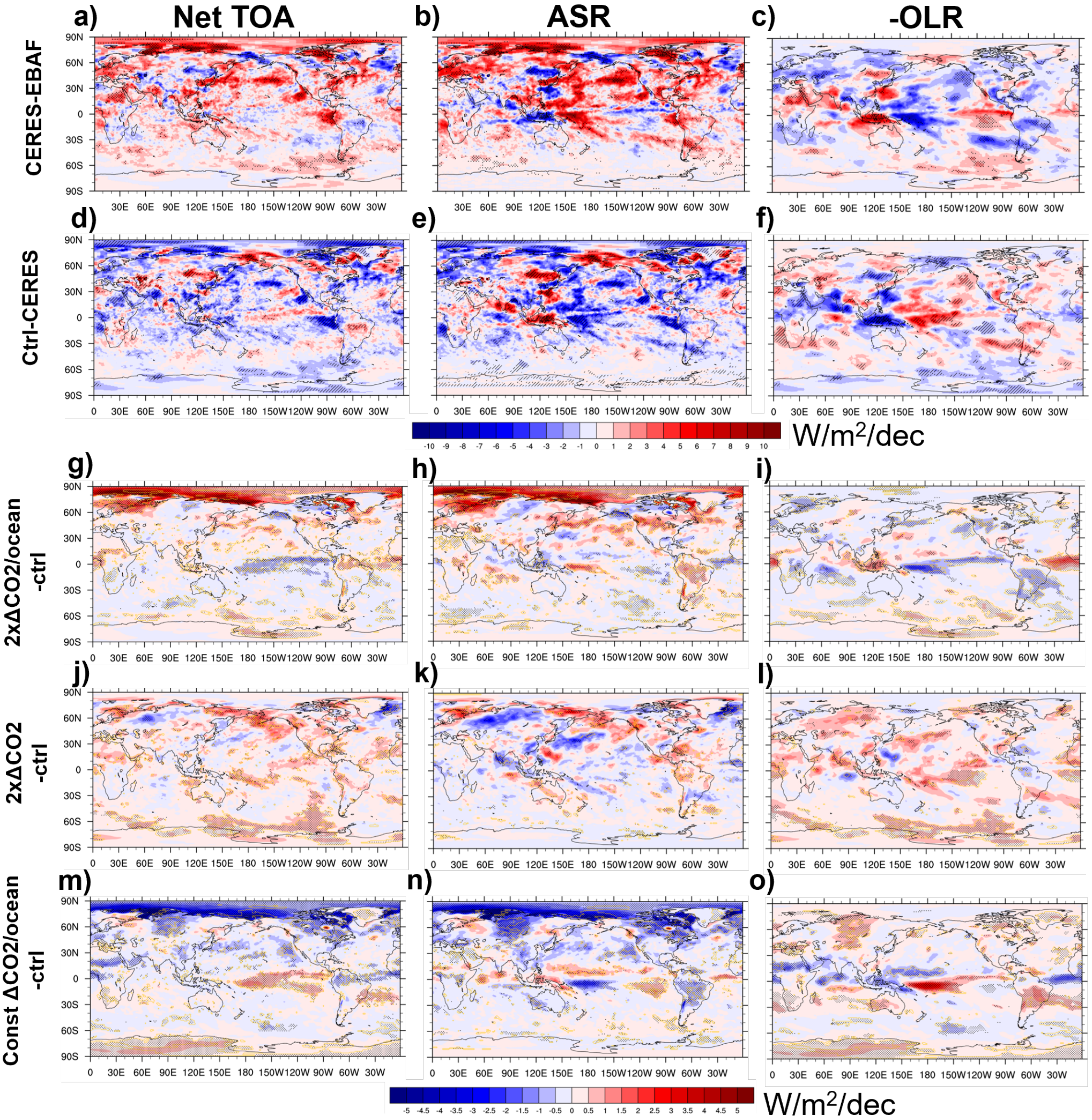

**Fig. 11. (left) Top-of-atmosphere net, (middle) absorbed shortwave, and (right) (-1*) outgoing longwave radiation trends for JJA 2001-2023: a)-c) show observed trends based on CERES-EBAF, d)-f) show trend errors of the CTRL hindcast, and the other rows show the trend differences between different experiments and the CTRL. In the first row, stippling indicates statistically significant (first row) trends. In the second row, hatching indicates where observed trends lie outside the central 90% of the bootstrapped trend distribution of the CTRL hindcast. From the third row onward, stippling indicates significant trend differences (90% confidence level). The orange**

**isolines indicate where trend differences mean an improvement of trends compared to the CTRL hindcast.**

The Double $\Delta CO_2$ experiment (Figs. 11j–l) isolates the impact of enhanced atmospheric $CO_2$ forcing alone. In this case, ASR trend changes are much smaller than in Double $\Delta CO_2$/ocean (Fig. 11h and Fig. 10b), while changes in −OLR dominate the net TOA response. The experiment therefore largely reflects the nearly instantaneous radiative effect of increased $CO_2$ concentrations in the absence of the Planck feedback.

The TOA trend changes in Constant $CO_2$/ocean (Figs. 11m–o) are broadly the mirror image of those in Double $\Delta CO_2$/ocean. Degraded ASR and net TOA trends occur over the tropical Pacific, North Pacific and the Arctic, resulting in global-mean trend changes that closely oppose those of Double $\Delta CO_2$/ocean (Fig. 10 and Table S1).

Finally, Table S1 also summarizes the effects of aerosol perturbations. Double ΔAER exhibits slightly enhanced global-mean ASR trends, driven by increased ASR trends over the oceans but reduced trends over land (Figure S7). The latter appears to result from unrealistically strong semi-direct aerosol effects in IFS (Tim Stockdale, personal communication). Changes in −OLR partially offset the ASR response, leaving only a modest net TOA trend impact in Double ΔAER. Consistent with this behaviour, Double $\Delta CO_2$/ocean/AER produces TOA trends that are very similar to those of Double $\Delta CO_2$/ocean, both globally (Table S1) and regionally (Figure S7).

### 6.5. Trends in the tropics

This section assesses the sensitivity of simulated tropical Pacific trends to the imposed forcing and IC perturbations.

Observed SST trends (Fig. 12a) exhibit the well-documented strengthening of the equatorial Pacific zonal SST gradient, accompanied by strengthening easterly trade winds in the central equatorial Pacific (Fig. 12b) and increasing precipitation over the Indo-Pacific warm pool (Fig. 12c). The trend toward a more La Niña-like mean state is also reflected in a northward displacement of the Pacific ITCZ. Trends in 200-hPa velocity potential show increasing values over much of the central Pacific extending to the Southern subtropical Atlantic (the western hemisphere) and decreasing values extending from western Africa to the Indo-Pacific Warm Pool (the eastern hemisphere). The resulting enhancement of the upper tropospheric zonal velocity-potential gradient is consistent with a strengthening Walker circulation.

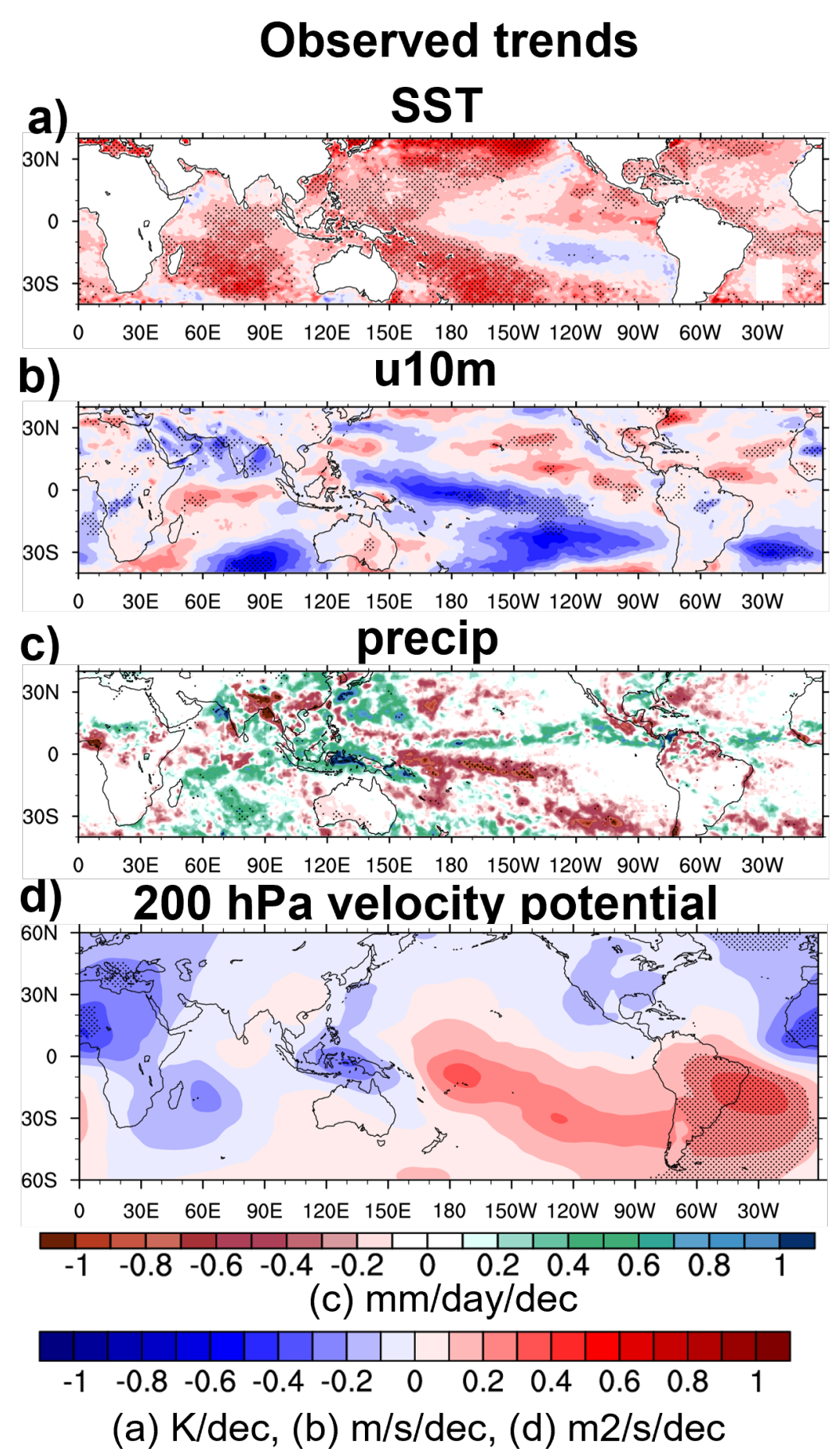

**Fig. 12. Observed JJA 1993-2023 trends of (a) SST (ERA5), (b) u10m (ERA5), (c) precipitation (GPCP), and (d) 200hPa velocity potential (ERA5). Stippling indicates statistically significant trends (90% confidence level).**

The CTRL hindcast reproduces many aspects of the observed SST trend pattern (compare Figs. 12a and 13b) but fails to capture the observed behaviour in the equatorial Pacific, where it exhibits spurious warming trends (see also Fig. 14b). This deficiency was already identified in the operational SEAS5 system (Mayer et al., 2025) and remains largely unchanged despite the updated model version and ocean ICs used here. As in SEAS5, the SST trend errors are accompanied by severe underestimation of the strengthening easterly wind trends in the central equatorial Pacific and precipitation trends over the Indo-Pacific warm pool (Figs. 13 and 14). Consequently, the hindcasts fail to reproduce the observed strengthening of the Walker circulation. The 200-hPa velocity-potential trend pattern shows little agreement with observations, with a pattern correlation of only 0.06 and approximately 32% of tropical grid points exhibiting observed trends outside the hindcast trend distribution (Figs. 13k and 14k).

Fig. 15 provides further insight into the evolution of these errors with lead time. Owing to the annual re-initialization, the zonal SST-gradient trend in May is well represented in CTRL. In contrast, the associated zonal wind and warm-pool precipitation trends are already substantially underestimated during the first forecast month. As the forecast evolves, the weakened atmospheric response erodes the SST-gradient trend, leading to a spurious Niño-3.4 warming trend and a weakened zonal SST gradient by JJA.

The Double $\Delta CO_2$/ocean experiment (right columns in Figs. 13 and 14) sheds light on the role of the forced ocean signal. Because the imposed perturbation amplifies the large-scale SST warming pattern, SST trends generally increase throughout the tropics. More importantly, however, the zonal SST-gradient trend contained in the ocean ICs is also substantially enhanced (roughly doubled). This stronger gradient is largely retained during the first forecast month, increasing the May SST-gradient trend from 0.09 ± 0.22 K/dec in CTRL to 0.18 ± 0.22 K/dec in Double $\Delta CO_2$/ocean (Fig. 15 and table S2), compared with an observed trend of 0.10 ± 0.20 K/dec.

The stronger SST gradient is accompanied by a stronger easterly wind trend. In May, the Niño-3.4 zonal wind trend approximately doubles, from −0.06 ± 0.18 m/s/dec in CTRL to −0.12 ± 0.18 m/s/dec in Double $\Delta CO_2$/ocean (Fig. 15 and table S2). Nevertheless, the simulated trend remains substantially weaker than the observed value of −0.26 ± 0.20 m/s/dec. By JJA, easterly wind trends are enhanced relative to CTRL (Figs. 13e,f), but the equatorial Pacific trend distribution remains inconsistent with observations (Fig. 14f), and the Niño-3.4 wind-trend error is reduced by only about 14% (table S2).

The limited atmospheric response results in a rapid erosion of the enhanced SST-gradient trend during the forecast. By JJA, the gradient trend decreases to 0.02 ± 0.29 K/dec in Double $\Delta$CO2/ocean, substantially below the observed value of 0.08 ± 0.32 K/dec (Fig. 15), even though the imposed ocean perturbation contains an approximately doubled gradient trend.

The impact on Niño-3.4 SST trends reflects the competition between two effects. On the one hand, the stronger zonal SST gradient enhances easterly winds and acts to suppress warming in the eastern Pacific. On the other hand, the imposed warming perturbation directly increases SST trends throughout the basin. The latter effect dominates, leading to a 33% increase in the spurious Niño-3.4 warming trend in Double ∆CO2/ocean relative to CTRL (table S2).

Warm-pool precipitation trends in JJA are also enhanced in Double $\Delta CO_2$/ocean and are improved by approximately 35% relative to CTRL (table S2), although they remain weaker than observed (Fig. 15 and Table S2). Despite the only modest improvements in precipitation and surface wind trends, the large-scale upper-tropospheric tropical circulation response is substantially improved. The 200-hPa velocity-potential trend pattern develops positive trends over the subtropical Pacific and South America and negative trends over western Africa, both in better agreement with observations (compare Figs. 12d and 13l). As a result, the pattern correlation increases from 0.06 in CTRL to 0.52, while the fraction of tropical grid points with observationally inconsistent trends decreases from 32% to 18% (Fig. 14l).

Trend changes in Constant $CO_2$/ocean (left columns of Figs. 13 and 14) are broadly opposite to those in Double $\Delta CO_2$/ocean and are therefore not discussed in detail. Two aspects are nevertheless noteworthy. First, substantial trends remain despite the removal of the estimated forced signal from the ocean ICs. Trends in Constant $CO_2$/ocean_ORAS6 are only slightly reduced compared to Constant $CO_2$/ocean (see Table S2), indicating that the remaining trends are robust to the period and data set to estimate the forced signal in the ocean. This suggests that trends associated with natural variability, changes in the observing system, or residual signals in other components of the ICs continue to contribute to trends in the hindcasts with constant forcing and the forced pattern removed from the ICs (see also Section 6.1). Second, the responses of zonal wind and SST-gradient trends are larger in magnitude than those in Double $\Delta CO_2$/ocean, whereas precipitation responses are comparatively weaker.

Finally, tropical Pacific trends in Double $\Delta CO_2$, in which only atmospheric $CO_2$ concentrations are modified, are nearly indistinguishable from those in CTRL. This indicates that direct radiative effects of increased $CO_2$ have little influence on tropical Pacific circulation trends on seasonal forecast timescales. Instead, the response is controlled primarily by the ocean ICs, which carry the accumulated effects of past radiative forcing.

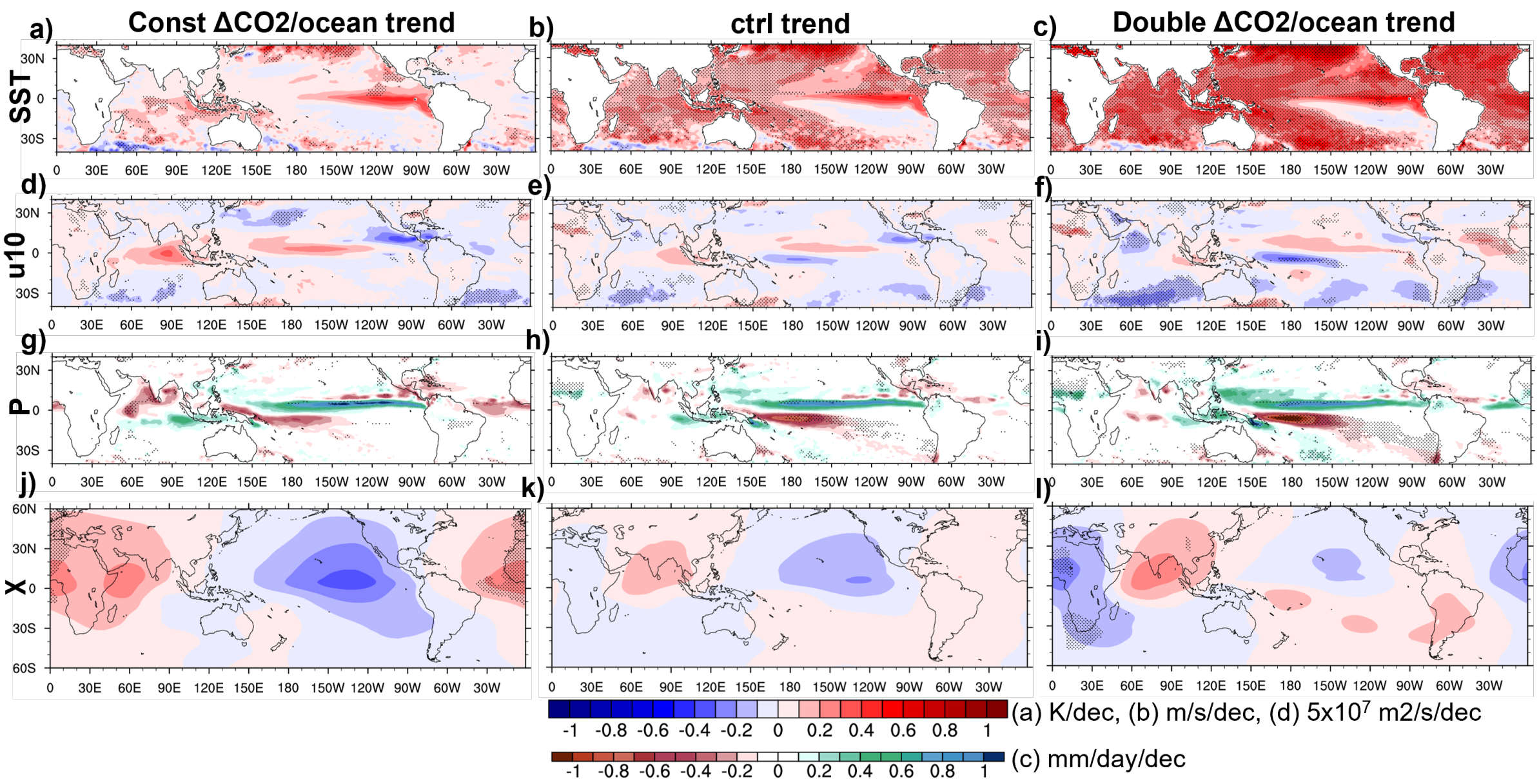

**Fig. 13. Ensemble mean hindcast trends for JJA 1993-2023 of (first row) SST, (second row) u10m, (third row) precipitation, and (fourth row) 200hPa velocity potential, based on (left column) Constant CO2/ocean, (middle column) CTRL, and (right column) Double ∆CO2/ocean. Stippling indicates significant trends on the 90% confidence level.**

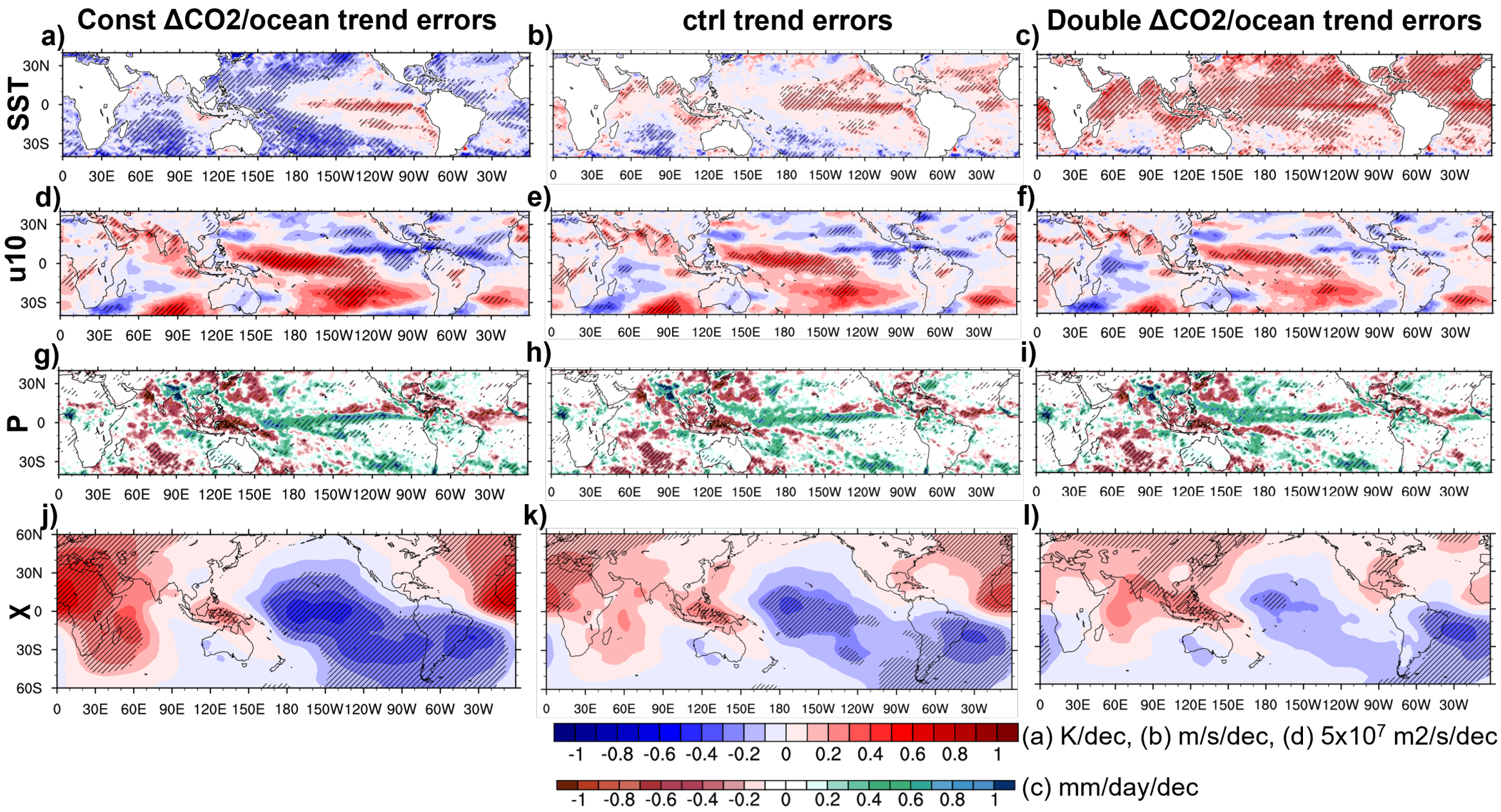

**Fig. 14. As Fig. 13, but for trend errors. Reference is ERA5 except for GPCP which serves as reference for precipitation. Hatching indicates where observed trends lie outside the inner 90% of the bootstrapped trend distribution of the hindcast.**

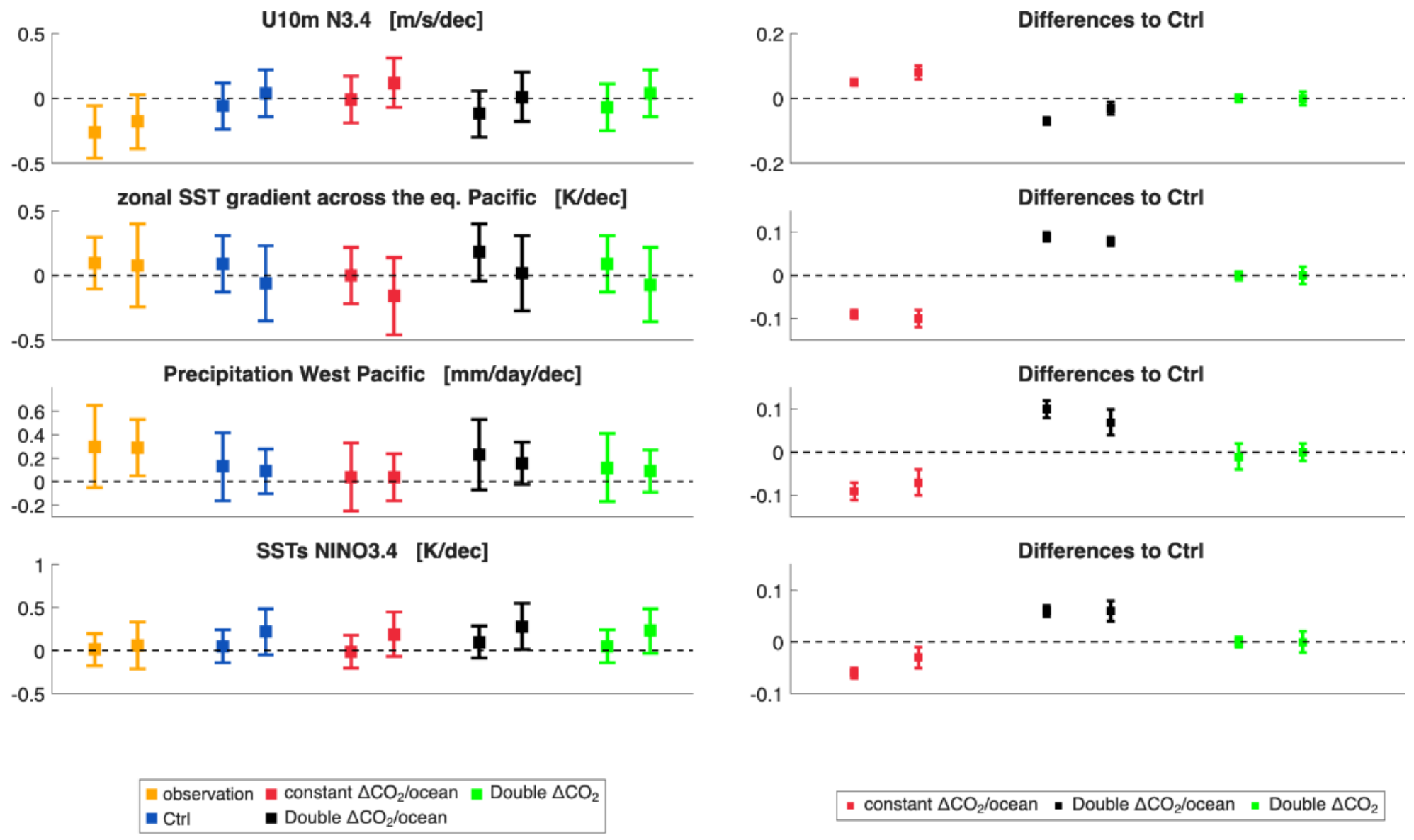

**Fig. 15. Trends in tropical Pacific key climate metrics for May (the left of each pair of values) and JJA (the right of each pair of values) 1993-2023. The metrics are 10-metre zonal wind (u10m) in Nino 3.4, zonal SST gradient (defined as difference of SSTs in 5S-5N, 110-180E and 5S-5N, 180-80W), precipitation in Western Pacific Warm pool (20S-20N, 120-150E), and SST in Nino 3.4. Panels on the left show ensemble mean trends and their uncertainties (whiskers indicate 90% confidence intervals estimated using a bootstrapping approach), and panels on the right show the trend differences with the CTRL. Trends for other experiments and more statistics are shown in Table S2.**

## 7. Summary and conclusions

This paper presents the experimental design and initial evaluation of a set of counterfactual seasonal hindcasts using the ECMWF coupled forecasting system. While perturbations to atmospheric $CO_2$ and aerosol forcing are straightforward to implement, the development of physically consistent ocean and sea-ice initial-condition (IC) perturbations require additional methodological considerations and testing. Because the hindcasts are initialized from the observed climate state, we chose to derive the IC perturbations from observational estimates of the forced climate response. This approach avoids introducing model-specific biases that may be present in free-running climate simulations such as those from CMIP. In constructing the forced signal estimate, a compromise was required between maximizing the length of the observational record and maintaining data quality. We therefore based the ocean-temperature forced pattern on the 1975–2023 observational record.

The resulting counterfactual hindcasts exhibit the expected response to altered forcing. Simulations with amplified forcing show enhanced long-term warming trends, while simulations in which forcing is held constant exhibit substantially reduced trends. At the same time, interannual variability and seasonal prediction skill are largely preserved, for example for ENSO. The counterfactual seasonal hindcasts experiments therefore achieve their primary objective: simulating the evolution of seasonal climate anomalies from observationally constrained initial states under alternative background climate conditions. This is a clear advantage over free-running model simulations for which it is impossible to compare specific events under different climates.

A key requirement for such experiments is that the imposed perturbations do not generate artificial model adjustments that obscure the physical response of interest. We find that this requirement is largely satisfied. Changes in model drift are generally small relative to the baseline drift of the CTRL simulation, indicating that the imposed ocean and sea-ice perturbations are dynamically consistent with the model climatology. In particular, SST drift changes are much smaller in Double $\Delta CO_2$/ocean than in Double $\Delta CO_2$, where only the atmospheric forcing is modified. This demonstrates that jointly perturbing radiative forcing and ocean ICs produces a substantially better-balanced response than modifying atmospheric forcing alone. Consequently, the counterfactual hindcasts are not dominated by spurious drift and provide a suitable framework for further scientific investigation.

Several aspects of the simulated decadal climate trends are improved in Double $\Delta CO_2$/ocean relative to CTRL. Most notably, global net TOA radiation trends become more realistic because absorbed shortwave radiation (ASR) trends strengthen in regions where they are underestimated in the control simulation. The improvement arises through enhanced cloud and circulation responses rather than through direct radiative trapping by $CO_2$ alone, suggesting that the model's radiative response to observed climate change is too weak in the baseline configuration.

Trends in the tropical Pacific response also improve with amplified forcing. Double $\Delta CO_2$/ocean produces stronger easterly wind trends in the central equatorial Pacific, enhanced precipitation trends over the Indo-Pacific warm pool, and substantially more realistic upper-tropospheric velocity-potential trends indicative of a strengthening Walker cell. These improvements are associated with the strengthened zonal SST-gradient trend imposed through the ocean ICs.

At the same time, the experiments reveal an important limitation of the forecasting system. The primary issue is the non-persistence of the imposed SST-gradient trend itself. Indeed, the enhanced zonal SST-gradient trend introduced through the ocean ICs is largely retained during the first forecast month and produces a corresponding strengthening of the atmospheric response. However, this response remains much weaker than observed. As a result, the coupled atmosphere–ocean system is unable to maintain the enhanced SST gradient through Bjerknes feedbacks, leading to a gradual erosion of the imposed gradient during the forecast. By JJA, the simulated zonal SST-gradient trend has weakened substantially and remains below the observed trend despite being initialized with a considerably stronger gradient trend signal.

This behaviour points to a more fundamental deficiency in the model representation of tropical Pacific coupling. The atmosphere responds too weakly to changes in the zonal SST gradient, producing insufficient strengthening of equatorial easterlies, western Pacific convection, and the Walker circulation. The resulting weak atmospheric feedback limits the persistence of the SST-gradient anomaly and contributes to the continued spurious warming trend in the Niño-3.4 region. This interpretation is consistent with Mayer et al. (2025), who found that equatorial Pacific wind trends remain strongly underestimated even in atmosphere-only simulations forced with observed SSTs. Together, these results suggest that biases in atmospheric sensitivity to tropical Pacific SST gradients, rather than deficiencies in the representation of trends in the ocean initial conditions, are a key factor limiting the ability of current seasonal forecasting systems to reproduce observed tropical Pacific climate trends.

Future work will use these counterfactual hindcasts to investigate the extent to which the changed tropical Pacific and radiative trends influence regional climate trends and climate extremes outside the tropics. More broadly, the experiments demonstrate that counterfactual seasonal hindcasts provide a practical framework for separating the effects of changing background climate conditions from internal variability while remaining closely constrained by observations.

**Code and data availability:**

The IFS source code is available subject to a licence agreement with ECMWF. ECMWF member-state weather services and approved partners have granted access. The IFS code without modules for data assimilation is also available. IFS is available under an openIFS licence (http://www.ecmwf.int/en/research/projects/openifs, last access: 23 June 2026) for educational and academic purposes (currently version 48R1). The NEMO4.0 source code version used here is available at https://doi.org/10.5281/zenodo.5566313 (NEMO System Team 2021; Madec and the NEMO System Team 2024). The interfaces to NEMO can be obtained from ECMWF on request under the license described above.

Data from all experiments presented in this experiment will be made publicly available via ECMWF's MARS archive before publication.

CERES-EBAF data has been retrieved from https://ceres.larc.nasa.gov/data/, GPCP v3.2 data from https://data.nasa.gov/dataset/gpcp-precipitation-level-3-monthly-0-5-degree-v3-2-gpcpmon-at-ges-disc-5d4c3, and ERA5 data is available from the Copernicus Climate Change Service (C3S) Climate Data Store (CDS): https://cds.climate.copernicus.eu/datasets/reanalysis-era5-complete?tab=overview

**Author contribution**

MM prepared main experimental setup, curated the required input data, carried out the presented diagnostics, and led writing of the manuscript. All co-authors contributed to experimental design, interpretation of results, and development of the manuscript. DJB provided additional support with experimentation. AW acquired the funding.

**Competing interests**

The authors declare that they have no conflict of interest.

**Acknowledgments**

The authors thank Nicholas Leach (University of Oxford) for help with forced pattern detection, Chris Roberts (ECMWF) for help with the gradient-preserving ocean adjustments, and Charles Pelletier (ECMWF) for support with sea ice modifications. The authors are also grateful for scientific discussions with Jonathan Day, Retish Senan, and Tim Stockdale (all ECMWF). This work was financially supported by the EXPECT project, which has received funding from the European Union's Horizon Europe Framework Programme under Grant Agreement 101137656.

**References**

Allan, R. P., and C. J. Merchant, 2025: Reconciling Earth's growing energy imbalance with ocean warming. *Environ. Res. Lett.*, **20**, 044002. doi: 10.1088/1748-9326/adb448

Batté, L., I. Välisuo, M. Chevallier, J. C. Acosta Navarro, P. Ortega, and D. Smith, 2020: Summer predictions of Arctic sea ice edge in multi-model seasonal re-forecasts. *Clim. Dyn.*, **54**, 5013–5029. doi: 10.1007/s00382-020-05273-8

Beverley, J. D., M. Newman, and A. Hoell, 2024: Climate model trend errors are evident in seasonal forecasts at short leads. *NPJ Clim. Atmospheric Sci.*, **7**, 285, https://doi.org/10.1038/s41612-024-00832-w.

Browne, P., E. de Boisseson, S. Keeley, C. Pelletier, and H. Zuo, 2026: Sea ice data assimilation in ORAS6. *The Cryosphere*, **20**, 3299–3311, https://doi.org/10.5194/tc-20-3299-2026.

Byrne, H., R. Seager, and J. E. Smerdon, 2025: CMIP6 models cannot capture long-term forced changes in the tropical Pacific sea surface temperature gradient. *Nat. Commun.*, https://doi.org/10.1038/s41467-025-66839-w.

Cheng, L., and Coauthors, 2024: IAPv4 ocean temperature and ocean heat content gridded dataset. *Earth Syst Sci Data*, **16**, 3517–3546, https://doi.org/10.5194/essd-16-3517-2024.

Chrust, M., A. T. Weaver, P. Browne, H. Zuo, and M. A. Balmaseda, 2025: Impact of ensemble-based hybrid background-error covariances in ECMWF's next-generation ocean reanalysis system. *Q. J. R. Meteorol. Soc.*, **151**, e4914, https://doi.org/10.1002/qj.4914.

Fasullo, J. T., and Coauthors, 2024: An overview of the E3SM version 2 large ensemble and comparison to other E3SM and CESM large ensembles. *Earth Syst. Dyn.*, **15**, 367–386, https://doi.org/10.5194/esd-15-367-2024.

Forbes, R., A. Geer, K. Lonitz, and M. Ahlgrimm, 2016: Reducing systematic errors in cold-air outbreaks. *ECMWF Newsl.*, **146**, 17–22. doi: 10.21957/s41h7q7l

Gillett, N. P., and Coauthors, 2016: The detection and attribution model intercomparison project (DAMIP v1. 0) contribution to CMIP6. *Geosci. Model Dev.*, **9**, 3685–3697, https://doi.org/10.5194/gmd-9-3685-2016.

Good, S. A., M. J. Martin, and N. A. Rayner, 2013: EN4: Quality controlled ocean temperature and salinity profiles and monthly objective analyses with uncertainty estimates. *J. Geophys. Res. Oceans*, **118**, 6704–6716, https://doi.org/doi:10.1002/2013JC009067.

Haustein, K., M. R. Allen, P. M. Forster, F. E. L. Otto, D. M. Mitchell, H. D. Matthews, and D. J. Frame, 2017: A real-time global warming index. *Sci. Rep.*, **7**, 15417, https://doi.org/10.1038/s41598-017-14828-5.

He, J., and B. J. Soden, 2016: The impact of SST biases on projections of anthropogenic climate change: A greater role for atmosphere-only models? *Geophys. Res. Lett.*, **43**, 7745–7750, https://doi.org/10.1002/2016GL069803.

Hersbach, H., and Coauthors, 2020: The ERA5 global reanalysis. *Q. J. R. Meteorol. Soc.*, **146**, 1999–2049, https://doi.org/10.1002/qj.3803.

Huffman, G. J., R. F. Adler, A. Behrangi, D. T. Bolvin, E. J. Nelkin, G. Gu, and M. R. Ehsani, 2023: The new version 3.2 Global Precipitation Climatology Project (GPCP) monthly and daily precipitation products. *J. Clim.*, **36**, 7635–7655, https://doi.org/10.1175/JCLI-D-23-0123.1.

Johnson, S. J., and Coauthors, 2019: SEAS5: the new ECMWF seasonal forecast system. *Geosci. Model Dev.*, **12**, 1087–1117, https://doi.org/10.5194/gmd-12-1087-2019.

Leach, N. J., and Coauthors, 2021a: FaIRv2. 0.0: a generalized impulse response model for climate uncertainty and future scenario exploration. *Geosci. Model Dev.*, **14**, 3007–3036, https://doi.org/10.5194/gmd-14-3007-2021.

——, A. Weisheimer, M. R. Allen, and T. Palmer, 2021b: Forecast-based attribution of a winter heatwave within the limit of predictability. *Proc. Natl. Acad. Sci.*, **118**, e2112087118, https://doi.org/10.1073/pnas.2112087118.

——, and Coauthors, 2024: Heatwave attribution based on reliable operational weather forecasts. *Nat. Commun.*, **15**, 4530, https://doi.org/10.1038/s41467-024-48280-7.

Li, J.-L. F., and Coauthors, 2021: Improved ice content, radiation, precipitation and low-level circulation over the tropical pacific from ECMWF ERA-interim to ERA5. *Environ. Res. Commun.*, **3**, 081006, https://doi.org/10.1088/2515-7620/ac1bfe.

Loeb, N. G., and Coauthors, 2018: Clouds and the earth’s radiant energy system (CERES) energy balanced and filled (EBAF) top-of-atmosphere (TOA) edition-4.0 data product. *J. Clim.*, **31**, 895–918, https://doi.org/10.1175/JCLI-D-17-0208.1.

——, S.-H. Ham, R. P. Allan, T. J. Thorsen, B. Meyssignac, S. Kato, G. C. Johnson, and J. M. Lyman, 2024: Observational assessment of changes in Earth’s energy imbalance since 2000. *Surv. Geophys.*, **45**, 1757–1783, https://doi.org/10.1007/s10712-024-09838-8.

Madec and the NEMO System Team, 2024: *NEMO Ocean Engine Reference Manual*, https://doi.org/10.5281/zenodo.1464816.

Mayer, M., M. A. Balmaseda, F. Vitart, and S. Tietsche, 2025: Tropical Pacific Trends in the ECMWF Seasonal System and Implications for Predictions of the 2020–22 Triple-Dip La Niña. *J. Clim.*, **38**, 2989–3003, https://doi.org/10.1175/JCLI-D-24-0467.1.

Minobe, S., E. Behrens, K. L. Findell, N. G. Loeb, B. Meyssignac, and R. Sutton, 2025: Global and regional drivers for exceptional climate extremes in 2023-2024: beyond the new normal. *Npj Clim. Atmospheric Sci.*, **8**, 138, doi: 10.1038/s41612-025-00996-z

Morice, C. P., and Coauthors, 2021: An updated assessment of near-surface temperature change from 1850: The HadCRUT5 data set. *J. Geophys. Res. Atmospheres*, **126**, e2019JD032361, https://doi.org/10.1029/2019JD032361.

Muller, C. J., and P. A. O’Gorman, 2011: An energetic perspective on the regional response of precipitation to climate change. *Nat. Clim. Change*, **1**, 266–271, https://doi.org/10.1038/nclimate1169.

NEMO System Team, 2021: NEMO release-4.0 (release 4.0), https://doi.org/10.5281/zenodo.5566313.

Ortega, E., M. Vancoppenolle, C. Rousset, and E. Lemaire, 2025: New representation of sea ice salt dynamics within NEMO-SI3: Evaluation and impacts. *ESS Open Arch.*, https://doi.org/10.22541/essoar.175882978.85357195/v1.

Patterson, M., D. J. Befort, J. F. Lockwood, J. Slattery, and A. Weisheimer, 2025: The representation of surface temperature trends in c3s seasonal forecast systems. *Atmospheric Sci. Lett.*, **26**, e1316, https://doi.org/10.1002/asl.1316.

Rugenstein, M., S. Dhame, D. Olonscheck, R. J. Wills, M. Watanabe, and R. Seager, 2023: Connecting the SST pattern problem and the hot model problem. *Geophys. Res. Lett.*, **50**, e2023GL105488. doi: 10.1029/2023GL105488

Schmidt, G., 2024: Why 2023's heat anomaly is worrying scientists. *Nature*, **627**, https://doi.org/10.1038/d41586-024-00816-z.

Seager, R., N. Henderson, and M. Cane, 2022: Persistent discrepancies between observed and modeled trends in the tropical Pacific Ocean. *J. Clim.*, **35**, 4571–4584, https://doi.org/10.1175/JCLI-D-21-0648.1.

Simpson, I. R., and Coauthors, 2025: Confronting earth system model trends with observations. *Sci. Adv.*, **11**, eadt8035, https://doi.org/10.1126/sciadv.adt8035.

Smith, D. M., and Coauthors, 2022: Attribution of multi-annual to decadal changes in the climate system: The Large Ensemble Single Forcing Model Intercomparison Project (LESFMIP). *Front. Clim.*, **4**, 955414, https://doi.org/10.3389/fclim.2022.955414.

Stockdale, T. N., R. Senan, and R. Bilbao, 2022: *Harmonized CAMS and CMIP6 datasets for aerosols (CONFESS project report)*, https://confess-h2020.eu/wp-content/uploads/2022/12/confess-d2-1-v1-0.pdf. no doi available

Von Schuckmann, K., and Coauthors, 2023: Heat stored in the Earth system 1960–2020: where does the energy go? *Earth Syst. Sci. Data*, **15**, 1675–1709, https://doi.org/10.5194/essd-15-1675-2023.

Weisheimer, A., T. N. Palmer, N. J. Leach, M. R. Allen, C. D. Roberts, and M. A. Abid, 2025: CO2-induced climate change assessment for the extreme 2022 Pakistan rainfall using seasonal forecasts. *Npj Clim. Atmospheric Sci.*, **8**, 262, https://doi.org/10.1038/s41612-025-01136-3.

Wills, R. C., and Coauthors, 2026: Forced Component Estimation Statistical Method Intercomparison Project (ForceSMIP). *J. Clim.*, e250326, https://doi.org/10.1175/JCLI-D-25-0326.1.

Zhang, L., Y. Chen, K. B. Karnauskas, C. Wang, M. Collins, and X. Luo, 2025: The 2023/24 El Niño event exhibited unusually weak extratropical teleconnections. *Commun. Earth Environ.*, **6**, 595, https://doi.org/10.1038/s43247-025-02584-8.

Zuo, H., and Coauthors, 2024: ECMWF's next ensemble reanalysis system for ocean and sea ice: ORAS6. *ECMWF Newsl.*, **180**. doi: 10.21957/hzd5y821lk

## Supporting Information

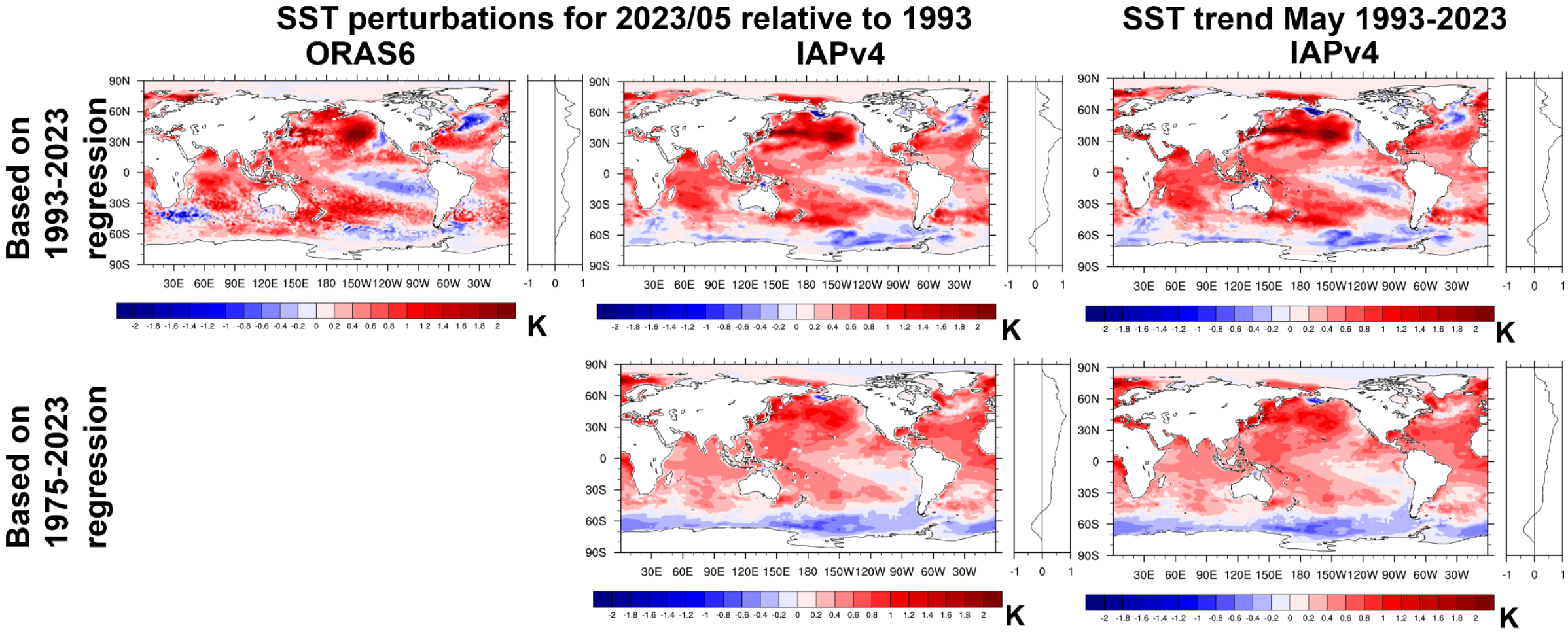

**Figure S1. Comparison of SST forced pattern estimates based on 1993-2023 (top left) ORAS6 and (top middle) IAP data to (top right) 1993-2023 SST trend based on IAP data. The second row compares (bottom middle) the forced SST pattern based on IAP 1975-2023 data to the (bottom right) 1975-2023 SST trend based on IAP data.**

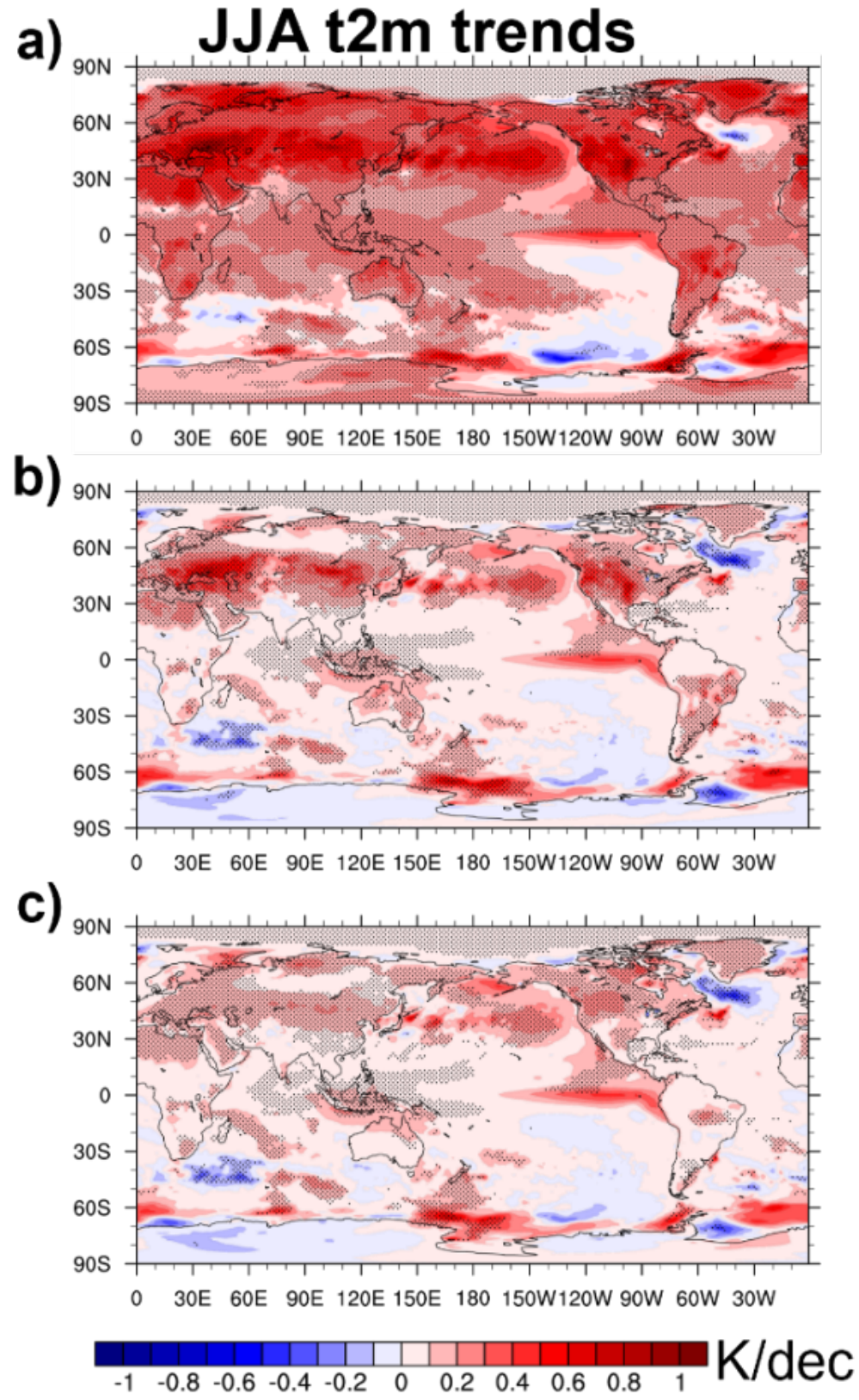

**Figure S2. T2M trends for JJA 1993-2023 in (a) CTRL, (b) Constant CO2/ocean, and (c) Constand CO2/ocean/ClimLand. Stippling indicates trends at 90% confidence level.**

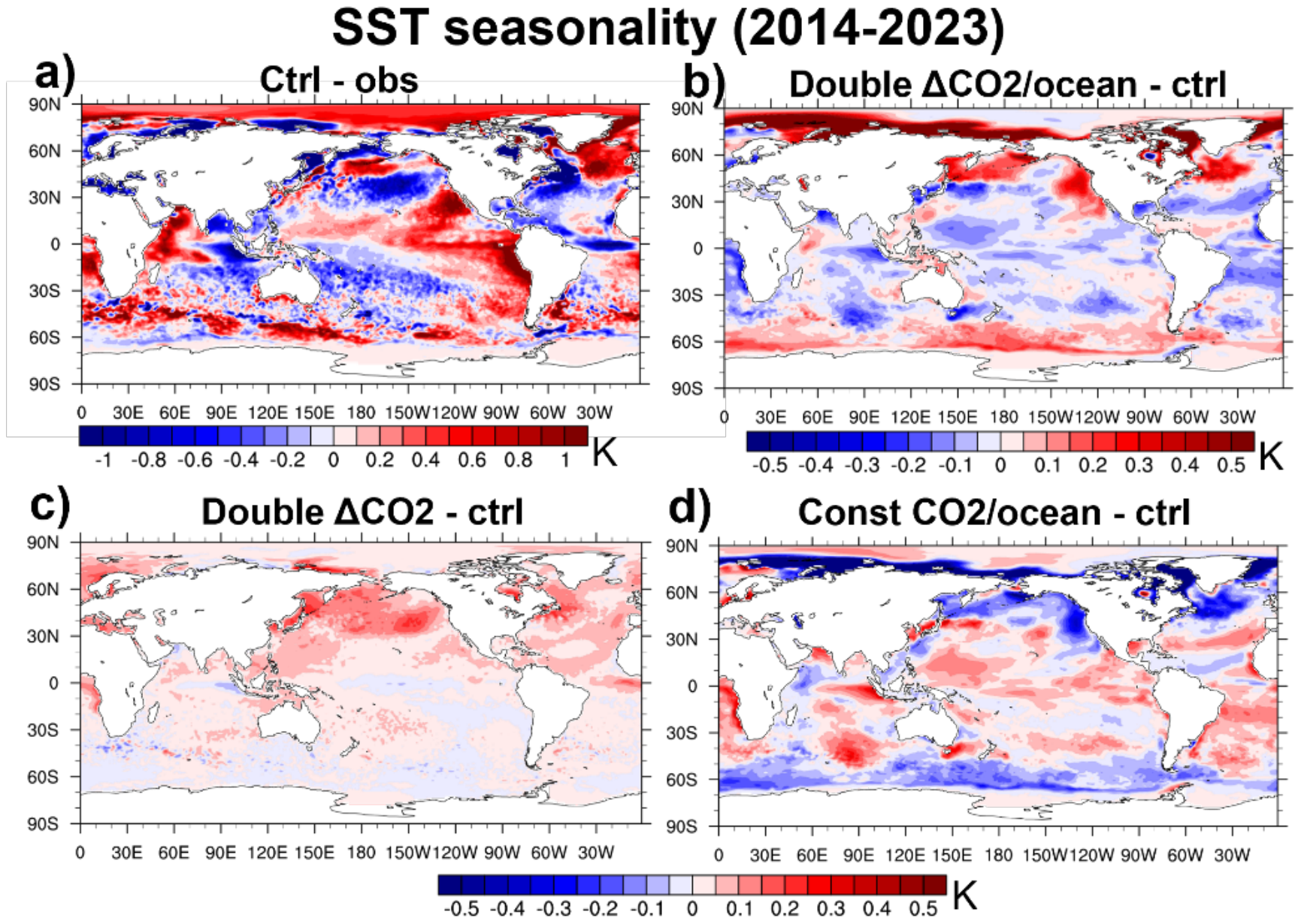

**Figure S3. (a) seasonal SST drift (May to August) in CTRL relative to observations (2014-2023); change in SST drift in (b) Double ∆CO2/ocean, and (c) Double ∆CO2, and (d) Constant CO2/ocean compared to CTRL.**

Figure S3a shows that the CTRL hindcast exhibits pronounced seasonal SST drift with a rich spatial structure. The baseline drift is not further explored here as we are mainly interested in the sensitivity of the model to the applied perturbations. The change in SST drift is considerably smaller than the baseline drift of the CTRL, and it appears mirrored between Double ∆CO2/ocean (Fig. S3b) and Constant CO2/ocean (Fig. S3d). In Double ∆CO2/ocean, we find a seasonal SST decrease relative to CTRL away from the polar regions. The likely cause is that the ocean loses heat to the atmosphere (note that atmospheric ICs remain unadjusted in our experiments and hence are cool compared to the perturbed ocean) during the first few weeks of the forecast. Indeed, the cooling pattern between roughly 40S and 40N is most pronounced during the first month of the forecast (not shown). Double ∆CO2/ocean exhibits strong relative warming of SSTs in the Arctic. This is plausible given that the Double ∆CO2/ocean ICs have negative Arctic sea ice perturbations which act to increase heat absorption during summer. Part of the increase of the seasonal cycle in the northern high latitude may also stem from the increased radiative forcing, the effect of which is strongest in the summer hemisphere where mixed layer depths are small, e.g. in the northern extratropical Pacific (see Fig. S3c). It is less obvious why Double ∆CO2/ocean exhibits seasonal warming relative to CTRL (i.e., weaker seasonal cooling given this is austral winter) in the Southern Ocean. Note the forced SST (sea ice) signal in the ICs is negative (positive) in that region (see Fig. 4a). We speculate that the negative SST perturbations may be seasonally damped by meridional atmospheric transports and changes to air-sea fluxes, effectively transferring the warm tropical SST perturbations to the Southern Ocean. There may also be a role for the changes in the 10m winds (see Fig. 6e). However, given that the wind changes are rather non-symmetric between Double $\Delta CO_2$/ocean and Constant $CO_2$/ocean (Fig. 6f) but SST changes are quite symmetric in the Southern Ocean contradicts a strong role for direct wind forcing of the seasonal SST changes in that region.

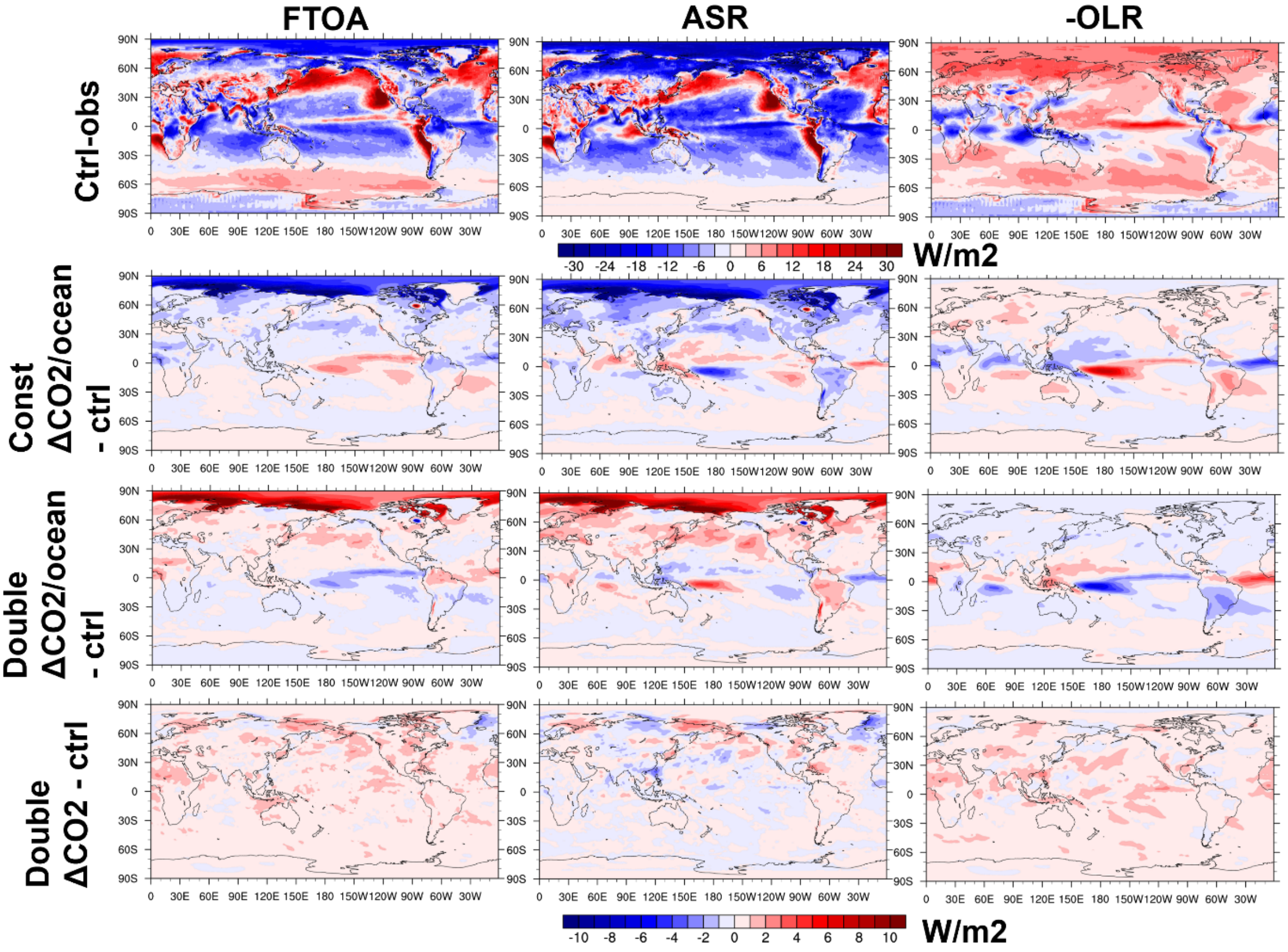

**Figure S4. Top-of-atmosphere (left) net, (middle) absorbed shortwave, and (right) (-1*) outgoing longwave radiation biases for JJA 2001-2023: first row shows bias of the CTRL simulation (CERES-EBAF used as reference), and the subsequent rows show the change in bias in a selection of experiments relative to CTRL.**

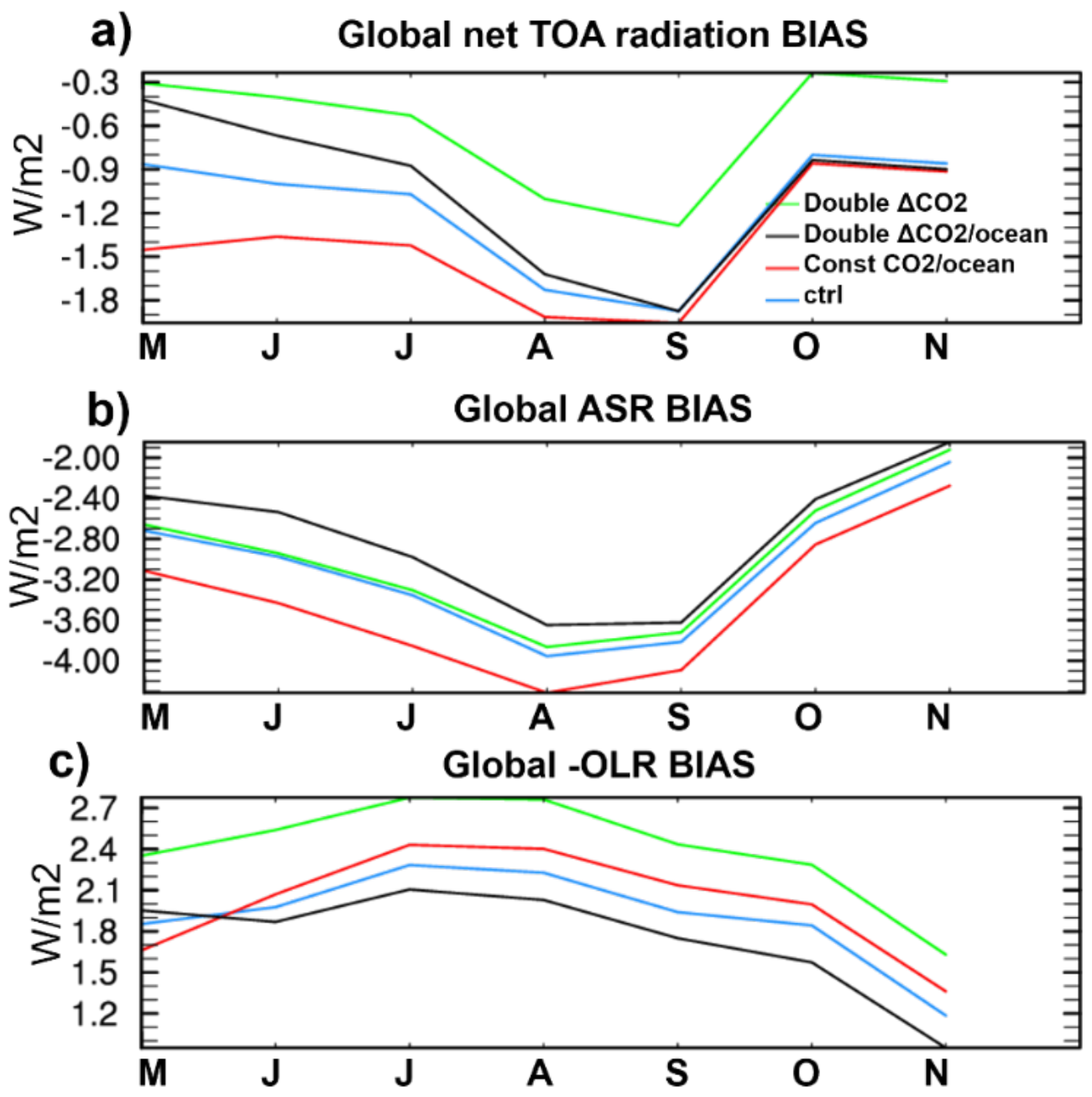

**Figure S5. Global mean top-of-atmosphere radiation biases for JJA 2004-2023: (a) net radiation, (b) ASR, (c) -OLR. CERES-EBAF is used as reference.**

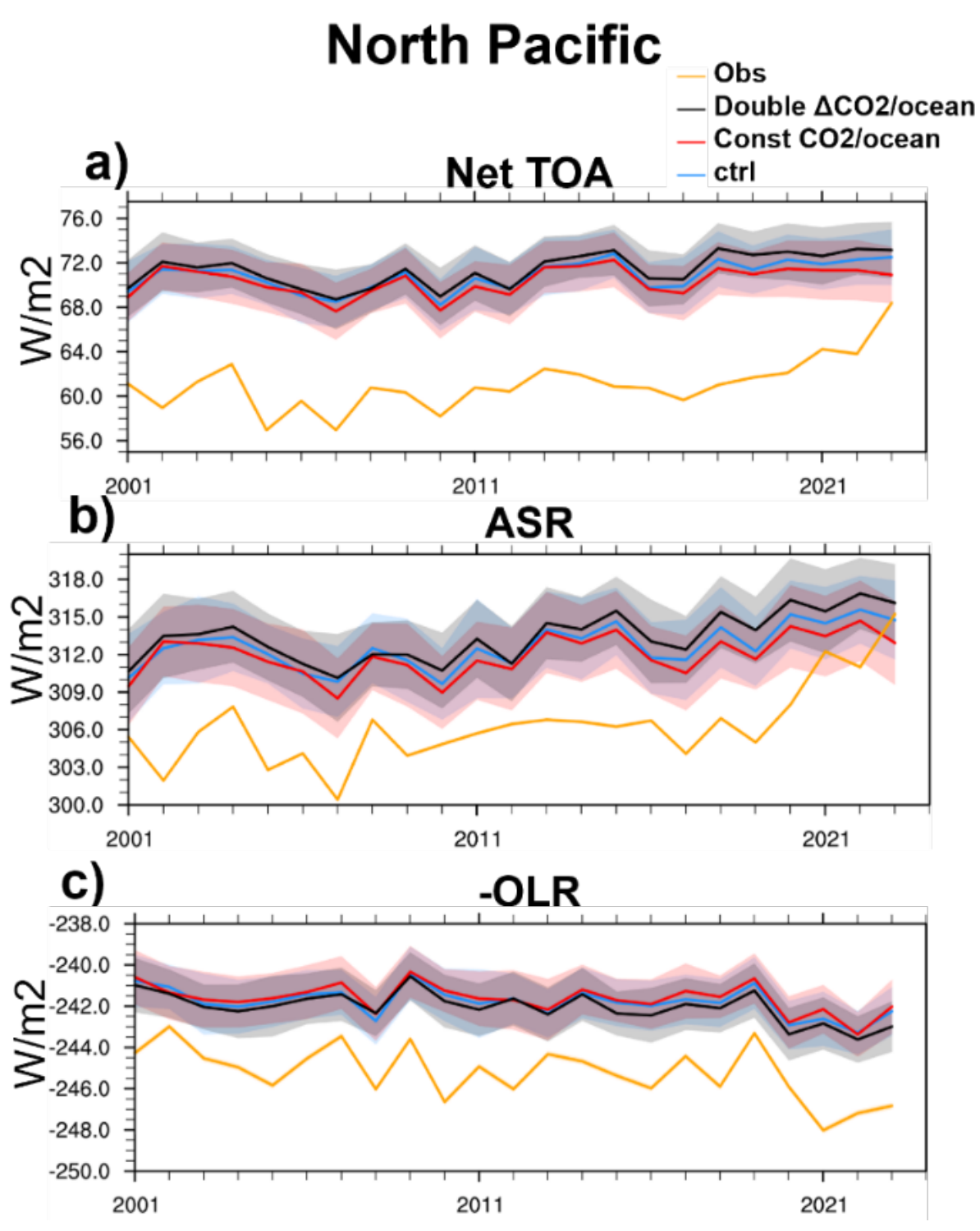

**Figure S6. North Pacific (30-55N, 140E-120W) average top-of-atmosphere radiation evolution for JJA 2001-2023: (a) net radiation, (b) ASR, (c) -OLR. CERES-EBAF is used as reference.**

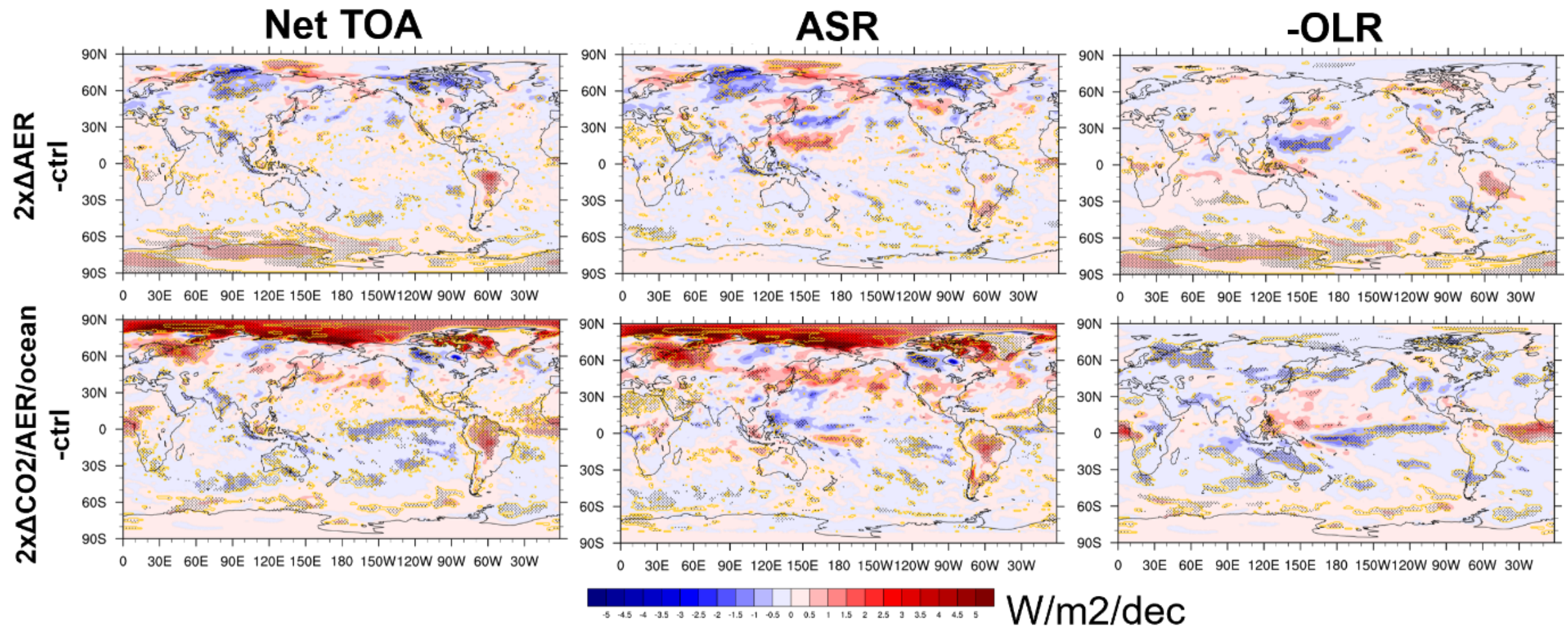

**Figure S7. As Fig. 11 in the main text, but for trend changes of (top row) Double ∆AER and (bottom row) Double ∆CO2/ocean/AER relative to the CTRL hindcast.**

| | Top-of-atmosphere radiation Trends JJA 2001-2023 [W/m2/decade] | | |
|---|---|---|---|
| | **Net** | **ASR** | **-OLR** |
| **CERES-EBAF** | **0.73 ± 0.23** | **0.86 ± 0.23** | **-0.13 ± 0.20** |
| **CTRL** | **0.25 ± 0.21** | **0.35 ± 0.24** | **-0.09 ± 0.15** |
| **Double ΔCO2/ocean** | **0.31 ± 0.21** | **0.47 ± 0.23** | **-0.16 ± 0.15** |
| **Double ΔCO2** | **0.48 ± 0.21** | **0.34 ± 0.23** | **+0.14 ± 0.14** |
| **Double ΔCO2/AER/ocean** | **0.32 ± 0.21** | **0.49 ± 0.24** | **-0.18 ± 0.16** |
| **Double ΔAER** | **0.29 ± 0.22** | **0.33 ± 0.23** | **-0.04 ± 0.14** |
| **Constant CO2/ocean** | **0.12 ± 0.21** | **0.15 ± 0.23** | **-0.03 ± 0.15** |
| **Double ΔCO2/ocean_ORAS6** | **0.35 ± 0.21** | **0.51 ± 0.24** | **-0.16 ± 0.15** |
| **Constant ΔCO2/ocean_ORAS6** | **0.08 ± 0.21** | **0.12 ± 0.24** | **-0.04 ± 0.14** |

**Table S1. Global mean top-of-atmosphere Trends JJA 2001-2023 [W/m2/decade] in observations and hindcast datasets. Hindcast trend central values are ensemble mean trends, and the 90% confidence intervals are obtained from the bootstrapped trend distributions.**

| | U10m N3.4 [m/s/dec] | | zonal SST gradient across the Pacific [K/dec] | | Precip West Pacific [mm/day/dec] | | Nino 3.4 [K/dec] | |
|---|---|---|---|---|---|---|---|---|
| | May | JJA | May | JJA | May | JJA | May | JJA |
| Obs | -0.26±0.20 | -0.18±0.21 | 0.10±0.20 | 0.08±0.32 | 0.30±0.35 | 0.29±0.24 | 0.01±0.19 | 0.06±0.27 |
| CTRL | -0.06±0.18* | 0.04±0.18* | 0.09±0.22 | -0.06±0.29* | 0.13±0.29* | 0.09±0.19* | 0.05±0.19 | 0.22±0.27* |
| Const ΔCO2 /ocean | -0.01±0.18* +25% | 0.12±0.19* +37% | 0.00±0.22* | -0.16±0.30*+71% | 0.04±0.29* +53% | 0.04±0.20* +30% | -0.01±0.19 -50% | 0.19±0.26*-24% |
| Const ΔCO2 /ocean minus CTRL | 0.05±0.01 | 0.08±0.02 | -0.09±0.01 | -0.10±0.02 | -0.09±0.02 | -0.07±0.03 | -0.06±0.01 | -0.03±0.02 |
| Const ΔCO2 /ocean_ORAS6 | -0.01±0.18* +25% | 0.10±0.17 +27% | 0.02±0.22* | -0.15±0.30*+64% | 0.02±0.29* +65% | 0.02±0.19* +35% | -0.03±0.19 0% | 0.12±0.27*-38% |
| Const $\Delta CO_2$/ocean_ORAS6 minus CTRL | 0.05±0.01 | 0.06±0.03 | -0.07±0.01 | -0.09±0.02 | -0.11±0.02 | -0.07±0.03 | -0.08±0.02 | -0.10±0.03 |
| Double $\Delta CO_2$/ocean | -0.12±0.18 -30% | 0.01±0.19*-14% | 0.18±0.22* | 0.02±0.29 -57% | 0.23±0.30 -58% | 0.16±0.18* -35% | 0.10±0.19* +125% | 0.28±0.27*+33% |
| Double $\Delta CO_2$/ocean minus CTRL | -0.07±0.01 | -0.03±0.02 | 0.09±0.01 | 0.07±0.01 | 0.10±0.02 | 0.07±0.03 | 0.06±0.01 | 0.06±0.02 |
| Double $\Delta CO_2$/ocean_ORAS6 | -0.10±0.19 -20% | 0.05±0.20 +5% | 0.16±0.22* | 0.01±0.31 -50% | 0.25±0.29 -71% | 0.16±0.18* -35% | 0.13±0.18* +300% | 0.37±0.27*+94% |
| Double $\Delta CO_2$/ocea | -0.04±0.02 | 0.01±0.03 | 0.07±0.01 | 0.07±0.02 | 0.13±0.03 | 0.07±0.03 | 0.08±0.02 | 0.15±0.03 |
| n_ORAS6 minus CTRL | | | | | | | | |
| Double $\Delta CO_2$ | -0.07±0.18* | 0.04±0.18* | 0.09±0.22 | -0.07±0.29 | 0.12±0.29* | 0.09±0.18* | 0.05±0.19 | 0.23±0.26* |
| Double $\Delta CO_2$ minus CTRL | 0.00±0.01 | -0.00±0.02 | 0.00±0.01 | 0.01±0.01 | -0.01±0.03 | -0.00±0.02 | 0.00±0.01 | 0.00±0.02 |

**Table S2. Trends in tropical Pacific key climate metrics for May and JJA 1993-2023. The metrics are 10-metre zonal wind (u10m) in Nino 3.4, zonal SST gradient (defined as difference of SSTs in 5S-5N, 110-180E and 5S-5N, 180-80W), precipitation in Western Pacific Warm pool (20S-20N, 120-150E), and SST in Nino 3.4. The asterisk indicates statistically significant trend differences with observations on the 90% confidence level. Statistically significant trend differences between counterfactual hindcasts and the CTRL are printed in bold. Red and green percentages indicate trend degradations and improvements compared to the CTRL, respectively.**